\documentclass{aa}  
\usepackage[utf8]{inputenc}
\usepackage{slashbox}
\usepackage{natbib}
\usepackage[colorlinks]{hyperref}
\usepackage{xcolor}
\usepackage{ulem}
\usepackage{booktabs}
\usepackage{array}
\usepackage{adjustbox}
\usepackage{arydshln}
\usepackage{placeins}
\usepackage{comment} 
\usepackage{caption}
\hypersetup{
    citecolor=[RGB]{44, 117, 255}, 
}
\usepackage{amsmath}
\usepackage{listings,comment}
\usepackage{graphicx}
\usepackage{subcaption} 
\def\tng{IllustrisTNG}

\def\TTH{\texttt{THE THREE HUNDRED }}
\makeatletter
\renewcommand*\aa@pageof{, page \thepage{} of \pageref*{LastPage}}
\makeatother

\title{Cosmic gas accretion from filaments onto galaxy clusters\\ using the IllustrisTNG simulation} 
\titlerunning{Cosmic accretion from filaments onto clusters}
\author{
  Jade Pasté \inst{1,2},
  Céline Gouin \inst{2},
  Nabila Aghanim \inst{1},
  Jenny G. Sorce \inst{3,1}
}

\institute{
Université Paris-Saclay, CNRS, Institut d’Astrophysique Spatiale, 91405 Orsay, France
\label{inst1}
\and
Sorbonne Université, UMR7095, Institut d’Astrophysique de Paris, 75014 Paris, France
\label{inst2}
\and
Université de Lille, CNRS, Centrale Lille, UMR 9189 CRIStAL, F-59000 Lille, France
\label{inst3}
}

\date{\today}
\abstract
{

Galaxy clusters grow through matter accretion from the cosmic web, mainly along filaments.
We aim to characterize gas accretion onto clusters, focusing on the role of filaments in driving anisotropic inflows and shaping thermodynamic properties, as it remains a key challenge in cosmology. 
In this study, we analyzed $415$ galaxy clusters from the IllustrisTNG-300 hydrodynamical simulation at $z=0$. 
Anisotropic signatures are highlighted by probing -- both isotropically and anisotropically (gas in filaments only) -- the radial profiles of gas properties (including temperature, entropy, density, and pressure) and radial velocity distributions.
Our results reveal two distinct regimes of gas accretion depending on the cluster-centric distances.
In the cluster environment ($\sim 2$–$4 R_{200}$), fast infalling warm gas channeled by cosmic filaments enters the warm-hot circumcluster medium, but filaments remain colder due to their slow thermalization with the surrounding medium, generating transverse temperature gradients. 
At the cluster outskirts ($\sim 1$–$2R_{200}$), gas infalling along filaments enters the hot intracluster medium (ICM), with a strong tangential velocity gradient. Warm gas penetrates clusters from filaments, while hot gas is preferentially ejected beyond them.
The mass and dynamical state of clusters significantly impact these accretion features, with relaxed and massive clusters exhibiting stronger and more extended temperature discontinuities. 
Overall, this work emphasizes a coherent picture of anisotropic gas accretion from filaments onto clusters. While virial shocks are observed near the cluster boundary, in particular at the filament–cluster interface, we do not find strong evidence of accretion shocks around filaments. This suggests the slow thermalization of filament gas as it enters the dense warm-hot circumcluster environment.}

\keywords{Galaxies: cluster: general -- large-scale structure of Universe -- Methods: statistical -- Methods: numerical }

\authorrunning{Pasté et al.}

\begin{document}
\maketitle

\section{Introduction}

At large scales, matter in the Universe is structured into a network-like distribution known as the cosmic web \citep{bond,deLapparent1986}, composed of nodes, filaments, walls, and voids. This large-scale architecture results from the gravitational collapse of small density fluctuations in the early Universe, as predicted by structure formation theory \citep{Zeldovich1970,Doroshkevich1970,Bardeen1986} and supported by N-body dark matter (DM) simulations \citep{Pichon_2011,Codis2012,Shim2021}.
Underdense regions expand, while overdense regions deepen and grow. As a result, matter flows from voids to walls, from walls to filaments, and from filaments to nodes \citep{AragonCalvo2010}. 

In the context of large-scale structures, galaxy clusters represent high-density regions, located at the nodes of the cosmic web, where filaments intersect. As such, they trace the most massive and gravitationally bound structures in the Universe.
Consequently, galaxy clusters are expected to grow continuously through the accretion of matter along filaments \citep{springel2005}. This accretion process is inherently anisotropic, with matter preferentially flowing along specific directions.
Filaments contain DM, galaxies, and gas, although the latter is more diffuse and colder than the gas found in clusters \citep[e.g.,][]{cen1999,dave2001}.
Since a significant fraction of baryons is expected to reside in filaments in a diffuse warm phase that is challenging to detect observationally, recent studies (e.g., \citealt{Tanimura2020,degraaf2019}) have provided increasing evidence for its presence, contributing to our understanding of the so-called "missing baryons" problem. Observationally, the baryonic content of galaxy clusters is relatively well constrained through X-ray and Sunyaev–Zel’dovich observations of the hot intracluster medium (ICM) \citep[e.g.,][for reviews]{Allen2011,Walker_2019}. In contrast, cosmic filaments remain particularly challenging to detect due to their low gas densities (around $10^{-6}$ to $10^{-4} \mathrm{cm}^{-3}$) and intermediate temperatures, generally ranging between $10^5$ and $10^7\mathrm{K}$ \citep{dave2001, galarraga2020,galarraga2021,gouin2022}, whereas the hotter ICM is strongly luminous in hard X-rays.

As cosmic filaments connecting galaxy clusters play a crucial role in the physics of matter accretion, they remain difficult to probe observationally. Nevertheless, direct detections have been achieved in a few exceptional systems, such as Abell 2744 \citep[e.g.,][]{Gallo_2024} or the Shapley supercluster, where \citet{Migkas_2025} report the first X-ray spectroscopic detection of pure warm–hot intergalactic medium (WHIM) emission from an individual 7.2\,Mpc filament. Hydrodynamical simulations have proven instrumental in exploring the complex processes governing this accretion \citep{Gouin21,gouin2022,Gouin_soft_2023,galarraga2020}. While previous studies have primarily focused on the statistical properties of isolated filaments in the cosmic web (\citealt{galarraga2020,galarraga2021}), the interface between filaments and galaxy clusters, where accretion flows transition from large-scale structures into virialized environments, remains complex due to the mixing of different gas phases and velocity components \citep{Lebeau2025,Lebeau2026}.

Galaxy clusters are not isolated spherical systems in hydrostatic equilibrium but are continuously influenced by anisotropic matter accretion along filaments \citep{Gouin21,gouin2022}, which affects both their morphology and X-ray emission properties \citep{Gouin25}. 
Using the \TTH project \citep{Cui2017}, which consists of a suite of $\sim300$ zoom-in resimulations of galaxy clusters, \cite{Rost2021} demonstrated that gas preferentially enters clusters through filaments and is ejected exclusively in regions outside filamentary structures. This behavior arises from the gravitational channeling of gas along cosmic filaments toward nodes, where transverse pressure confinement sustains coherent inflows. Ram pressure and accretion shocks at the cluster boundary then expel gas preferentially into the lower-density inter-filament regions. \cite{Rost2024} further examined the thermodynamical properties and gas dynamics from cluster outskirts to large distances using zoom-in resimulated clusters from the same simulation suite, with the aim of identifying potential accretion shocks. However, both studies analyzed the gas properties in spherical shells around clusters, excluding the inner regions up to $2R_{200}$.
\cite{Vurm2023} employed the C-EAGLE simulation to study individual clusters from their centers to the far cosmic web to investigate the thermodynamical characteristics of cluster-connected filaments, clearly identifying gas inside versus outside filamentary structures.
Similarly, \cite{Lebeau2025} analyzed a simulated cluster to study the gas properties inside and outside filaments, highlighting the turbulence generated at the interface between the ICM and infalling filamentary gas.

The present study focuses on the mechanisms governing gas accretion from cosmic filaments into galaxy clusters. Our main goal is to build a comprehensive and coherent picture of how gas is accreted and thermally processed during its infall into clusters.
We used the IllustrisTNG300-1 simulation at $z=0$, which provides a large sample of clusters with a realistic description of the gas physics in their environments.
In contrast to previous studies based primarily on spherically averaged quantities, we explicitly highlight the anisotropic nature of gas accretion by directly comparing gas accreted through cosmic filaments with the surrounding circumcluster medium.
In this work, we link the gas thermodynamic properties to both (i) the filamentary geometry of accretion and (ii) cluster physical properties, such as mass and dynamical state. This allowed us to quantify how the cosmic web imprints the gas properties across different cluster populations and filament environments.

The paper is organized as follows. In Sect. \ref{Data}, we describe the sample of 415 clusters from the IllustrisTNG simulation along with the algorithm used to detect the filaments of the cosmic web. 
In Sect. \ref{Physics properties}, we examine the physical properties of the gas —such as temperature, density, pressure, and entropy— by analyzing these quantities for all clusters in the sample, and we also investigate the impact of the cluster relaxation state on gas dynamics.
In Sect. \ref{sect:Tan}, we further explore the transverse temperature profiles across filaments. 
In Sect. \ref{Gas dynamics}, we study the isotropic and anisotropic components of the radial gas velocity to highlight the anisotropic infall of gas toward the cluster.  Finally, we present our discussion and conclusions in Sect. \ref{sect:Discussion} and Sect. \ref{Conclusion}.

\section{Gas physics and filaments in IllustrisTNG}
\label{Data}

\subsection{IllustrisTNG simulation }
\label{Illustris}

The magnetohydrodynamic simulation \tng-300 \citep{TNG,TNG2} models the formation and evolution of DM, gas, and galaxies in a box of $302.6\,\rm Mpc$ from the early Universe to the present day ($z=0$). Gas dynamics and thermodynamics are computed on a moving adaptive mesh using the \texttt{AREPO} code \citep{springel2005}, which employs a finite-volume Godunov scheme on a quasi-Lagrangian mesh, enabling an accurate treatment of hydrodynamical fluxes across cell interfaces. The simulation adopts cosmological parameters from \cite{planck2016-cosmology}, including $\Omega_{\Lambda,0}=0.6911$, $\Omega_{m,0}=0.3089$, $\Omega_{b,0}=0.0486$, $\sigma_8=0.8159$, $n_s=0.9667$, and $h=0.6774$.

The DM mass resolution of TNG300-1 is $m_{\rm DM} = 4.0 \times 10^7 M_\odot / h$ for DM particles, and $m_{\rm gas} \sim 7.6 \times 10^6 M_\odot / h$ for gas cells. 
The gravitational softening length for collisionless components (DM and stars) is $\epsilon \sim 1\,{\rm kpc}/h$ (comoving), while the adaptive Voronoi mesh used for the gas reaches a minimum cell size of roughly $0.7\,{\rm kpc}$ in dense regions. The mesh adaptively refines and de-refines cells to keep their masses close to a target value, yielding higher spatial resolution in dense structures and coarser resolution in diffuse environments such as filaments.
This combination of box size and resolution enables the statistical analysis of cluster-connected large-scale structures, while resolving the anisotropic accretion of gas along cosmic filaments on sub-kiloparsec scales.

In this study, we used a sample of 415 galaxy clusters and groups from the halo catalogs of the IllustrisTNG simulation at $z=0$. IllustrisTNG halos are identified by the friends-of-friends (FoF) algorithm \citep{Davis1985}, applied to DM particles. Our cluster and group sample was selected from this FoF halo catalog by applying two conditions -- $M_{200}>5\times10^{13}M_{\odot}h^{-1}$ and halos that are more distant than $5R_{200}$ from the box edges \citep[similarly to][]{Gouin21} -- to ensure a well-resolved sample and avoid boundary effects. The radius, $R_{200}$, is defined as the radius of a sphere enclosing a mass, $M_{200}$, corresponding to a mean density equal to 200 times the critical density of the Universe.

The physical properties of the gas used in this work were extracted from the gas cells of the IllustrisTNG simulation. Starting from the electron abundance, $x_e$, (defined as the ratio of the free-electron number density to the total hydrogen number density $x_e \equiv n_e/n_{\rm H}$) and internal energy, $u$, we computed additional thermodynamic quantities, such as the mean molecular weight,
\begin{equation}
    \mu = \frac{4}{1 + 3 X_H + 4 X_H x_e} \cdot m_p \,,
    \label{eq:placeholder_label}
\end{equation}
and temperature, 
\begin{equation}
    T = (\gamma - 1) \ast \frac{u}{k_B} \ast \frac{\text{UnitEnergy}}{\text{UnitMass}} \ast \mu \,.
    \label{eq:temperature}
\end{equation}
Here, $\gamma=5/3$ denotes the adiabatic index, $k_b$ the Boltzmann constant, and $X_H=0.76$ the hydrogen mass fraction. Moreover, \text{UnitEnergy} and \text{UnitMass} denote the IllustrisTNG code units expressed in CGS units (in detail, $\mathrm{UnitEnergy}/\mathrm{UnitMass} = 10^{10}\,\mathrm{erg\,g^{-1}}$).

As for the entropy, we adopted the commonly used entropy proxy defined as $S = T \, n_e^{-2/3}$, with $n_e$ being the electron number density. This quantity is proportional to the specific thermodynamic entropy of an ideal gas and is conserved during adiabatic processes, while it is modified by nongravitational effects such as shocks, radiative cooling, and feedback processes. This definition assumes a fully ionized plasma in local thermodynamic equilibrium, an approximation that is well satisfied for the hot gas considered in this work.

\subsection{Cluster dynamical state}

To explore the impact of cosmic gas accretion on cluster properties, we estimated the dynamical state of our selected halo sample by evaluating their level of dynamical relaxation.  In practice, we determined their relaxation state according to two different proxies \citep[similar to][ for example]{Cui2017}.
A first proxy to estimate cluster dynamical state is the center-of-mass offset, computed as the distance between the center of mass, $r_{\rm cm}$, of the halo's main galaxy, and the density peak, $r_c$, that corresponds to the location of the minimum gravitational potential within the halo, normalized by the cluster radius, $R_{200}$:
\begin{equation}
\Delta r = \frac{|r_{\rm cm}-r_{\rm c}|}{R_{200}} .
\end{equation}

We also used the subhalo mass fraction, $f_{\rm sub}$, which is the fraction of mass contained in substructures within a halo:
\begin{equation}
f_{\rm sub} = \sum_i \frac{M_{\rm sub,i}}{M_{\rm tot}} .
\end{equation}
Here, $M_{\rm sub,i}$ denotes the mass of the $i$-th subhalo within the halo, and $M_{\rm tot}$ is the total mass, defined as the sum of the individual masses of all particles and cells of all types associated with the halo by the FoF algorithm.

Following these two definitions, we considered a halo to be relaxed if both conditions were satisfied, for example, with a low substructure fraction $f_{\rm sub}<0.1$ and a small center offset $\Delta r<0.1$. The threshold values were chosen according to the distribution of these quantities in our sample and appear to be the same as in the literature \citep[see, e.g.,][]{Cui2017}. In contrast, unrelaxed halos were selected to be both populated by massive substructures ($f_{\rm sub}>0.1$) and with a large offset between the center of mass and the density peak ($\Delta r>0.1$). 
Using these two proxies, we identify 95 relaxed clusters and 95 unrelaxed clusters out of the full sample of 415 halos described in Sect. \ref{Illustris}. The remaining systems satisfy only one of the two criteria and were therefore not included in either the relaxed or unrelaxed subsamples.

\subsection{Detection of cosmic filaments in IllustrisTNG}
\label{sect:disperse}

DisPerSE (Discrete Persistent Structure Extractor) \citep{Sousbie2011} is an algorithm designed to detect and extract multi-scale structures based on the identification and analysis of critical points in a density field.
Our methodology to reconstruct the cosmic web skeleton in IllustrisTNG is based on that used by \cite{Bahe2025}, who provide an optimal methodology to identify robust large-scale cosmic filaments from DM distribution in TNG simulations. Here, we summarize their methodology. First, they constructed a 3D smoothed DM density map, and second, they applied a DisPerSE filament finder algorithm on it.
We began by constructing the DM density field on a regular 3D grid of comoving cell resolution of $\sim0.4 {\rm Mpc}.h^{-1}$.
This grid resolution avoids the detection of intracluster filaments, given that our study focuses on gas accretion from large-scale filaments to clusters.
The DM density grid was then smoothed in logarithmic space to reduce the difference between low- and high-density regions \citep[as discussed by][]{Cautun2013}, with a spherical Gaussian kernel with a size of $0.5 {\rm Mpc}.h^{-1}$.

Finally, we applied the DisPerSE algorithm to the smoothed 3D DM density field, adopting an absolute persistence threshold of $T = 6000$ and a skeleton smoothing parameter of $S = 1$. These choices were motivated by the extensive analysis of \cite{Bahe2025} (see their Fig. 4), who explored the impact of these parameters on the TNG100 simulation. As discussed in Appendix \ref{appendixA}, the resulting filament skeleton consistently follows the underlying DM distribution.

\subsection{Substructure removal method}
\label{sect:substructure}

As demonstrated in \cite{galarraga2021}, the presence of substructures affects the thermodynamic profiles of gas in filaments. These substructures generally host warm, dense gas associated with galaxies and/or groups that are falling into clusters \citep{Gouin25}.
Here, we define subhalos as bound substructures identified within a FoF halo using the Subfind algorithm; they correspond to smaller halos orbiting within a larger cluster halo.
To exclude gas associated with substructures, we first identified all subhalos within $5R_{200}$ of each cluster center with masses exceeding $10^{11}M_{\odot}/h$ (excluding the main halo). We then computed the distance from each gas cell to its nearest subhalo. After testing various radial exclusion criteria, we adopted a threshold that removes gas within $2R_{\mathrm{halfmass}}$ (in comoving kiloparsecs per $h$) of each subhalo's center, where $R_{\mathrm{halfmass}}$ is the radius enclosing half of the total mass of the subhalo.

\section{Anisotropic imprints of filaments on gas thermodynamics \label{Physics properties}}

\begin{figure*}
    \centering

    \includegraphics[width=0.39\textwidth]{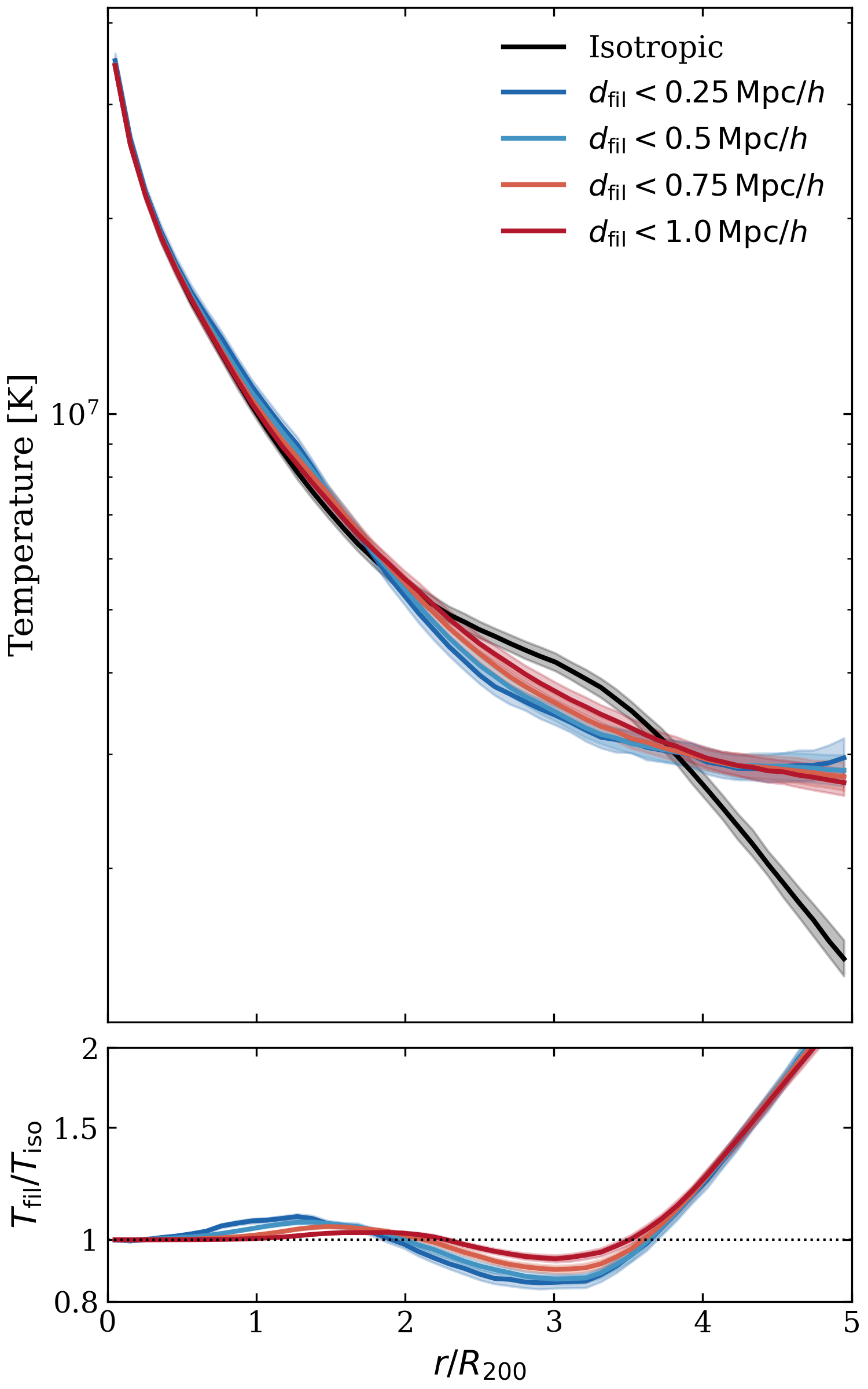}%
    \hspace{0.02\textwidth}%
    \includegraphics[width=0.39\textwidth]{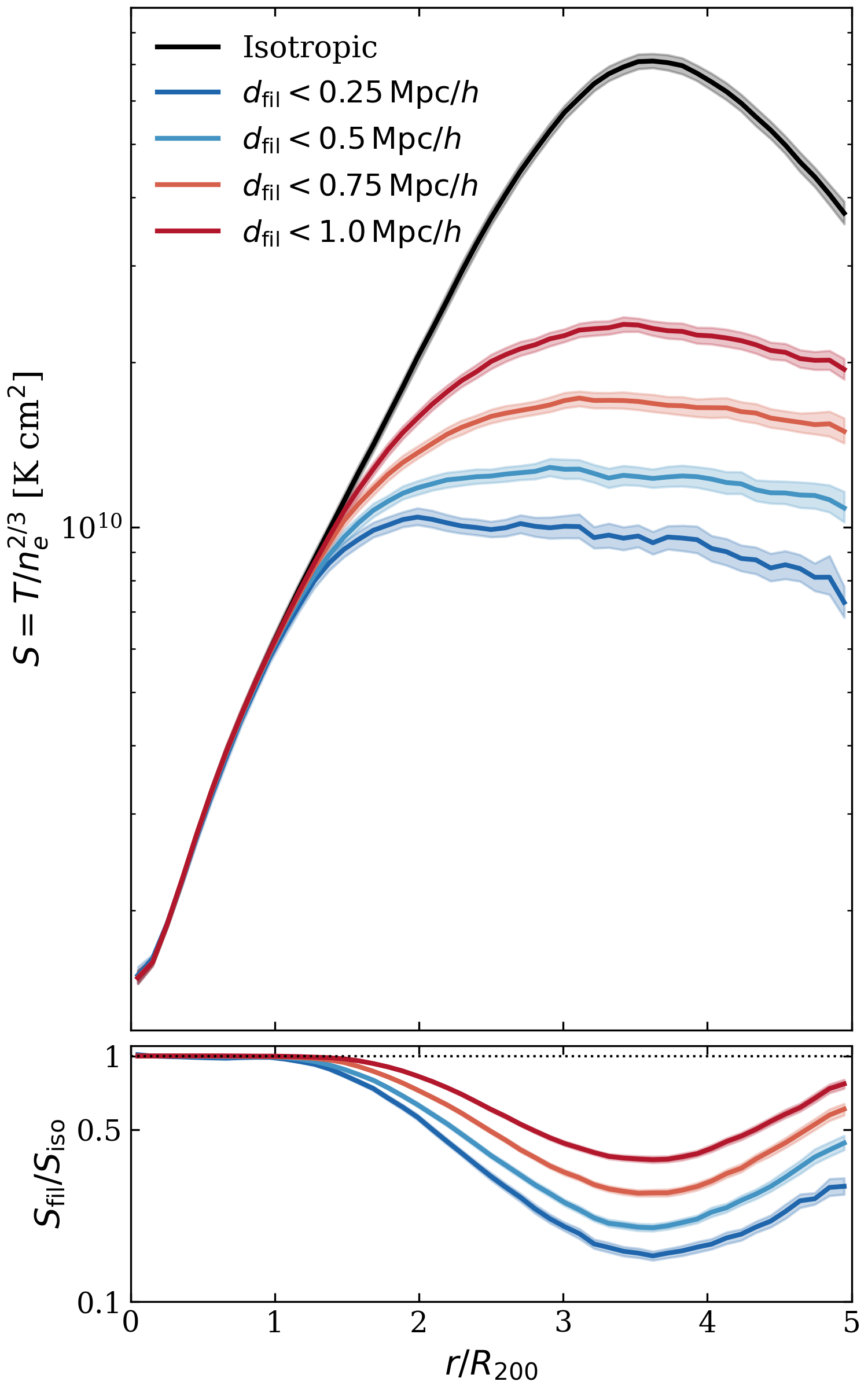}

    \vspace{0.1cm}

    \includegraphics[width=0.39\textwidth]{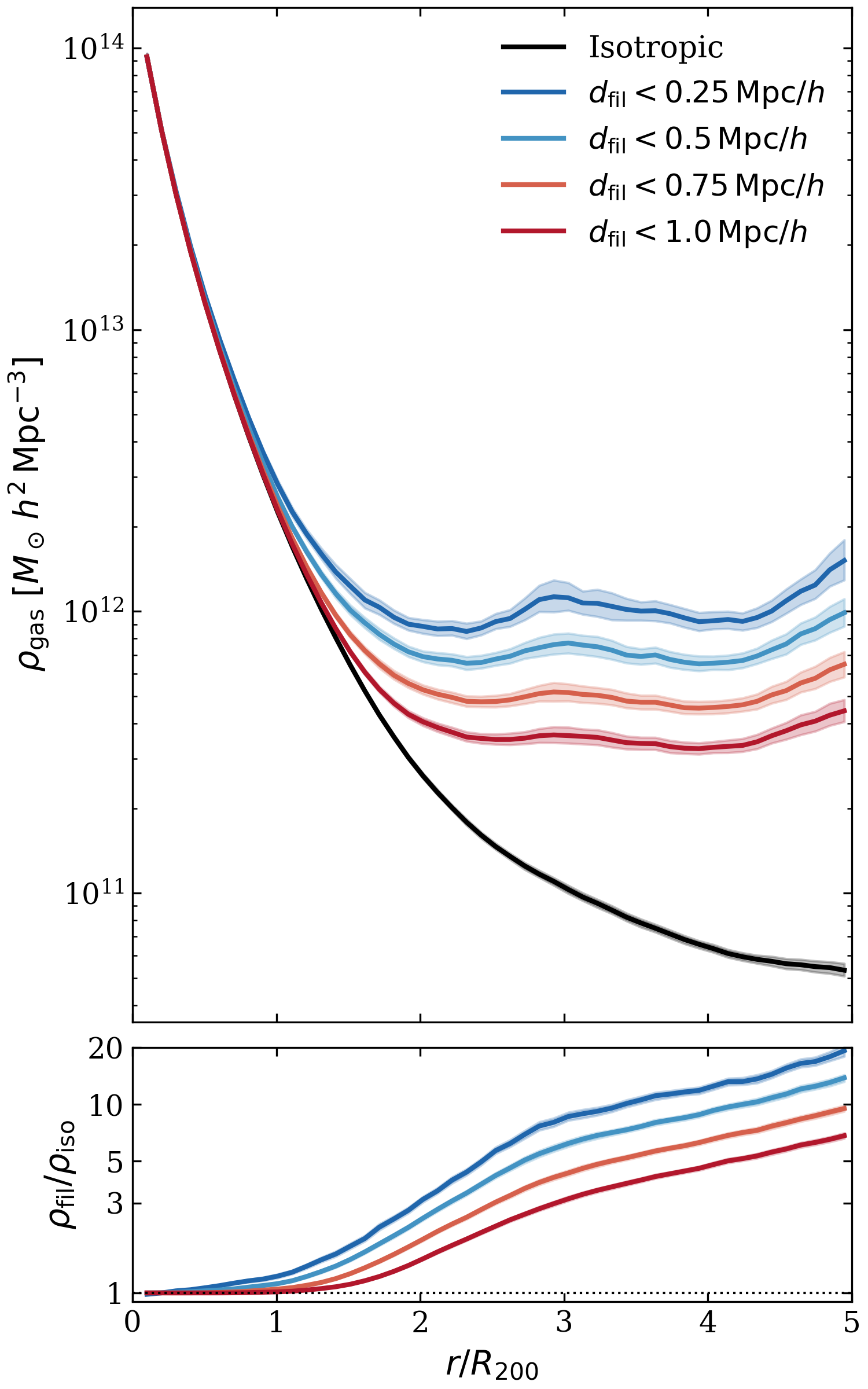}%
    \hspace{0.02\textwidth}%
    \includegraphics[width=0.39\textwidth]{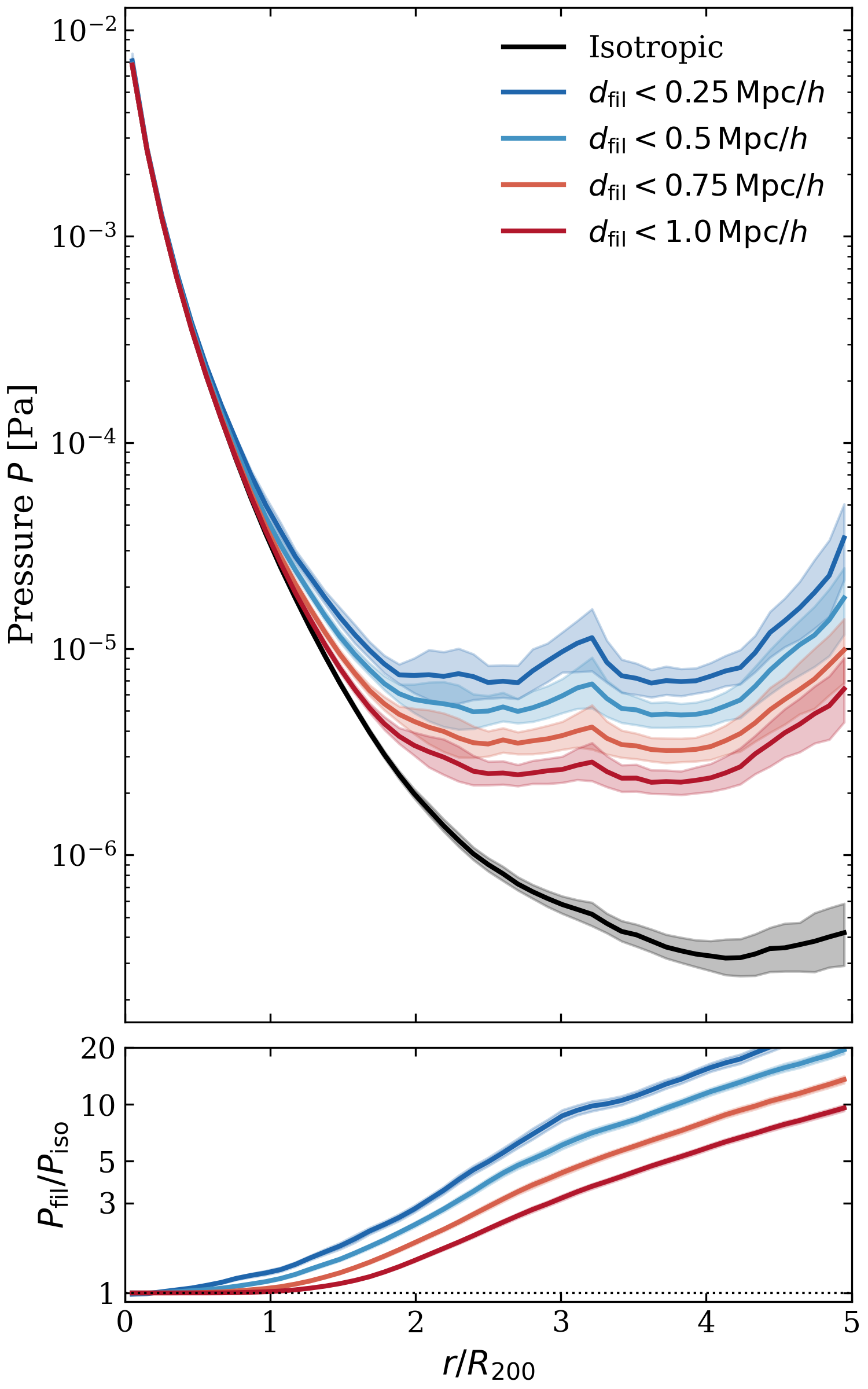}

    \caption{From top left to bottom right: Temperature, entropy, density, and pressure as a function of the distance from the cluster center, normalized by $R_{200}$. Top panels: Isotropic (black) and anisotropic (colored) profiles. Bottom panels: Ratio between the isotropic and anisotropic properties. The colors correspond to different filament thicknesses: $d_{\rm fil}\in[0.25,\,0.5,\,0.75,\,1]\,{\rm Mpc}.h^{-1}$. \label{fig:profile_properties} }
    \label{fig:profile_properties}
\end{figure*}

In this section, we aim to probe the influence of filaments on gas properties in cluster environments. While spherical averages of gas properties provide a global view, we show that this approach hides important directional effects induced by cosmic filaments, which channel accretion and shape the gas's thermodynamic state.
First, we aim to capture anisotropic signatures induced by cosmic filaments by probing the radial profiles of gas properties across our full cluster sample. Second, we analyze how these anisotropic imprints change depending on cluster properties.

\subsection{Methodology}

To highlight filament imprints, we probed the radial distribution
of gas properties as a function of their distance to clusters by considering two distinct definitions. First, isotropic profiles were obtained by spherically averaging gas properties in all directions. Second, anisotropic profiles were computed by taking into account only the gas in filaments.
In detail, gas was selected as a function of $d_{\rm fil}$, the perpendicular distance to the filament spine, by considering four thresholds of $d_{\rm fil}=0.25,0.5,0.75$ and $1 {\rm Mpc}.h^{-1}$.

Notice that we normalized the radial distance to clusters by the $R_{200}$ radius of each individual cluster. This allowed us to average radial profiles over clusters without mixing different physical scales intrinsic to different cluster sizes.

For each cluster, we computed the (isotropic and anisotropic) radial profile of a gas property $X$ as a function of cluster-centric distance $r$ using 50 bins from $0.05$ to $5R_{200}$. Within each bin, the gas property $X$ (temperature ($T$), entropy ($S$), or pressure ($P$)) is averaged over the gas cells using a volume-weighted average, where $V$ denotes the cell volume \citep[similar to][]{martizzi2019}:
\begin{equation}
\langle X \rangle (r) = \frac{\sum_{i \in r} X_{\rm i}  V_{\rm i}}{\sum_{i \in r} V_{\rm i}} .
\label{eq:Weight}
\end{equation}
We adopted a volume-weighted average as it is independent of the \texttt{AREPO} grid-- namely, of the varying volumes of the Voronoi gas cells \citep{Weinberger2020}. A comparison with mass-weighted estimates for temperature and entropy is provided in Appendix~\ref{AnnexeC}.

The gas density was similarly obtained as the total gas mass over the total gas volume in each radial bin:
\begin{equation}
\rho(r) = \frac{\sum_{i \in r} M_{\rm i}}{\sum_{i \in r} V_{\rm i}} .
\label{eq:rho}
\end{equation}

Finally, to highlight potential gas features along filaments, we further examined the ratio of anisotrapic to isotropic profiles, defined as $\text{Ratio} = X_{\rm filament} / X_{\rm isotropic}$.
\subsection{Anisotropic imprint of gas properties \label{sect: radial profiles}}

In Fig. \ref{fig:profile_properties}, we show the different radial profiles of gas thermodynamic properties, i.e., temperature, entropy, density, and pressure. The solid lines represent the mean profiles over our full cluster sample, with the error bars that are the $16$ and $84$-th percentile from bootstrap resampling. 
In each panel, the different colors indicate anisotropic radial profiles according to different distances to the filament $d_{\rm fil}$, whereas the black line indicates the isotropic radial profile. 
In general, we find that the ratio of anisotropic to isotropic profiles is close to unity below $0.75R_{200}$. This occurs because a large fraction of the cluster volume at these radii is occupied by filaments, which converge near the cluster center. The volume filling fraction 
of filaments reaches $90$--$100\%$ within $\sim 0.75$--$1\,R_{200}$ 
for the larger filament thicknesses considered here ($d_{\rm fil}\geq 0.75\,{\rm Mpc}.h^{-1}$). At these radii, the isotropic and filamentary directions probe nearly the same gas so their ratio naturally approaches unity.

The temperature profiles, shown in the top-left panel, reveal three distinct thermal regimes influenced by the filamentary structures in cluster environments. At large cluster-centric distances, $r \simeq [3.5 - 5]R_{200}$, filaments are hotter than the isotropic average, reaching temperatures up to twice as high. At this scale, there is no temperature gradient with $d_{\rm fil}$, suggesting that cosmic filaments are isothermal. This result is consistent with the findings of \cite{galarraga2021}, who explored the thermodynamics of the large-scale cosmic web. 
This is also consistent with expectations that filaments are primarily composed of warm gas with $T \sim 10^{5-7}\,\rm K$ \citep[see, e.g.,][]{tuominen2021}.
Interestingly, the isotropic and anisotropic temperatures converge at $\sim 3.5\,R_{200}$, suggesting that filaments enter the warm environment surrounding clusters. Between $3.5$ and $2\,R_{200}$, the radial temperature gradient is shallower along filaments than in their surroundings, resulting in temperatures up to $\sim 20\%$ lower around $3\,R_{200}$ compared to the ambient medium. This behavior likely reflects that filaments are not fully thermalized with the surrounding warm circumcluster medium, consistent with the higher infall velocities of the gas they channel (see Sect. \ref{Gas dynamics}). This is also suggested by the radial temperature gradient within filaments themselves: the filament center is colder than its outskirts, in agreement with \cite{Kotecha2022}. This indicates that, even though filaments are close to isothermal along the line connecting the cluster to the cosmic web at a given distance, they remain out of thermal equilibrium with the ambient medium they penetrate.

The trend reverses around $\sim 2R_{200}$, with filaments becoming hotter than the ICM and exhibiting an inverse tangential gradient (filament center hotter than outskirts). At the interface between filaments and clusters, the filament temperature exceeds the isotropic average by up to $\sim 10\%$, suggesting heating induced by the interaction between warm gas accreting along filaments and the hot ICM. This scenario is further developed throughout this paper. Notably, similar radial profiles inside and outside filaments have been observed around a cluster in the C-EAGLE simulation \citep{Vurm2023}.
Indeed, they identify a similar trend in the temperature profiles, observing a constant temperature inside filaments at large distances from the cluster center and noting that the gas outside filaments is hotter than that within filaments between $2$--$3R_{200}$.

Entropy behavior complements this picture. Outside clusters (beyond $R_{200}$), the entropy profiles rapidly increase by up to $\sim2 R_{200}$ for filaments and up to $\sim 3.5R_{200}$ for isotropic directions. 
Interestingly, anisotropic profiles show a lower entropy plateau, suggesting a continuous inflow of cooler, denser material and reduced mixing compared to the more turbulent cluster outskirts. Indeed, around $2R_{200}$—where the first temperature bump subsides— the entropy outside filaments becomes $0\%$ to $50\%$ higher than within the filaments (consistent with \cite{VallesPerez2021}).  

Finally, the density and pressure profiles exhibit a similar behavior. As expected, both density and pressure decrease rapidly with cluster-centric distance (similarly to \cite{Migkas_2025}). However, beyond $\sim 1 R_{200}$, gas located in filaments is denser and at higher pressure than gas along other isotropic directions. Moreover, a clear radial gradient is observed within filaments: their cores are denser and exhibit higher pressure compared to their outskirts. 
The combined analysis confirms that filaments act as distinct channels within the cluster environment. Gas inside filaments is denser, exhibits higher pressure, and has lower entropy, demonstrating a strong anisotropic imprint on the cosmic gas thermodynamics. This complex behavior also reveals two gradients in gas properties within $\sim 1$--$3.5\,R_{200}$: one radial gradient toward the cluster center and a tangential gradient from the filament spine to its outer regions, the latter weakening at larger cluster-centric distances where filaments approach thermal equilibrium with the surrounding medium.

\begin{table*}[t]
    \centering 
    \begin{tabular}{l@{\hspace{1.5cm}}c@{\hspace{1.5cm}}c@{\hspace{1.5cm}}c}
        \toprule
         & Relaxed & Unrelaxed & All \\
        \midrule
        
        Top 20\% mass $M\,[10^{13}M_\odot]$
        & \begin{tabular}{@{}c@{}}
        $N_{\rm cluster}=19$ \\
        $M\in[13.6,\,70.0]$\\
        $\langle M \rangle= 24.04$ 
        \end{tabular}
        & \begin{tabular}{@{}c@{}}
        $N_{\rm cluster}=19$ \\
        $M\in[13.4,\,88.6]$\\
        $\langle M \rangle= 28.59$ 
        \end{tabular}
        & \begin{tabular}{@{}c@{}}
        $N_{\rm cluster}=83$ \\
        $M\in[13.5,\,104.0]$\\
        $\langle M \rangle= 25.31$ 
        \end{tabular} \\
        \addlinespace  
        
        Bottom 20\% mass $M\,[10^{13}M_\odot]$
        & \begin{tabular}{@{}c@{}}
        $N_{\rm cluster}=19$ \\
        $M\in[5.05,\,5.81]$\\
        $\langle M \rangle= 5.355$
        \end{tabular}
        & \begin{tabular}{@{}c@{}}
        $N_{\rm cluster}=19$ \\
        $M\in[5.01,\,5.58]$\\
        $\langle M \rangle= 5.277$
        \end{tabular}
        & \begin{tabular}{@{}c@{}}
        $N_{\rm cluster}=83$ \\
        $M\in[5.01,\,5.83]$\\
        $\langle M \rangle= 5.346$
        \end{tabular} \\
        \addlinespace  
        
        All 
        & \begin{tabular}{@{}c@{}}
        $N_{\rm cluster}=95$ \\
        $M\in[5.05,\,70.0]$\\
        $\langle M \rangle= 11.09$ 
        \end{tabular}
        & \begin{tabular}{@{}c@{}}
        $N_{\rm cluster}=95$ \\
        $M\in[5.01,\,88.6]$ \\
        $\langle M \rangle= 11.71$
        \end{tabular}
        & \begin{tabular}{@{}c@{}}
        $N_{\rm cluster}=415$ \\
        $M\in[5.01,\,104.0]$\\
        $\langle M \rangle= 11.31$
        \end{tabular} \\ 
        
        \bottomrule  
    \end{tabular}    
    \caption{Mass ranges for the top and bottom 20\% of relaxed, unrelaxed, and all clusters. All masses are given in units of $10^{13}\,M_\odot$ (linear scale). $N_{\rm cluster}$ represents the cluster number considered in the corresponding sample.}
    \label{table:MassRange}
\end{table*}

\subsection{Impact of cluster properties on gas anisotropy \label{Relax&Mass}}

Here, we investigate the connection between gas anisotropy and cluster properties, focusing on halo mass and dynamical state (as defined in Sect.~\ref{Data}). A summary of our cluster sample and its sub-selection in mass and relaxation is provided in Table~\ref{table:MassRange}.
To assess the impact of mass, we selected the top and bottom 20\% of the mass distribution, ensuring comparable sample sizes. We also defined two cluster samples, a relaxed and an unrelaxed one, based on their dynamical state. By combining mass and relaxation criteria, we constructed four distinct subsamples of 20 clusters.

\begin{figure}
    \centering
    \includegraphics[width=1\linewidth]{FIGURES/all_ratios_three_panels_combined_415.png}
    \caption{Ratio of the anisotropic and isotropic temperature for different cluster selection: the $20\%$ most and least massive (top panel), relaxed and unrelaxed (middle panel), and the combination of relaxation and mass criterion (bottom panel). The cluster selection is summarized in Table~\ref{table:MassRange}.}
    \label{fig:ratio_relaxation_mass}
\end{figure}

Mass is a central property, as it sets the depth of the gravitational potential well. It thus governs the temperature and luminosity cluster scaling relation, as it directly affects the heating and compression of the ICM. Previous studies have indeed demonstrated its influence on various cluster characteristics, including dynamics \citep{Evrard2008}, luminosity \citep{Lovisari_2021}, thermal properties \citep{Pratt_2010,Sun2009}, and even the number of filaments connected to clusters \citep{Codis2012}.

The top panel of Fig.~\ref{fig:ratio_relaxation_mass} shows the ratio of temperature radial profiles between anisotropic and isotropic averages. A strong dependence on cluster mass is evident: the temperature drop inside filaments around clusters ($\sim 1.5 - 3.5R_{200}$) is significantly stronger for massive clusters, reaching up to 35\% below the isotropic temperature. 
In contrast, low-mass systems exhibit a similar trend but with minor temperature fluctuations between filaments and isotropic directions.
Notice that outside clusters' environments (beyond $4R_{200}$), filaments around massive clusters are hotter than those around less massive clusters, consistent with previous studies showing that filaments connected to denser environments exhibit higher temperatures \citep{galarraga2020,galarraga2021}.

In addition, we observe in the middle panel of Fig.~\ref{fig:ratio_relaxation_mass} that the cluster dynamical state has an impact on temperature variation. 
Indeed, inside cluster environments ($<4R_{200}$), only relaxed halos show a significant temperature discrepancy between filaments and the isotropic average.

Finally, cluster mass and dynamical state are intrinsically related, with massive halos generally less relaxed than low-mass ones \citep{Power_2011,Kuchner2020,Seppi_2021}. To explore their combined effect, we divided the sample into four subgroups (Table~\ref{table:MassRange}): massive relaxed, massive unrelaxed, low-mass relaxed, and low-mass unrelaxed, shown in the bottom panel of Fig.~\ref{fig:ratio_relaxation_mass}.
The halo mass strongly affects filament temperatures. Relaxed massive clusters exhibit the strongest temperature contrast, with filament gas being up to $\sim$40\% colder than the surrounding cluster outskirts at $3R_{200}$. High-mass unrelaxed clusters, however, show a slighter and less steep decrease, with the minimum located closer to the cluster center, around $2.5R_{200}$.
For low-mass relaxed systems, temperature variations remain modest, with filaments reaching up to 10\% higher temperatures around $2R_{200}$ and up to 10\% lower values around $3.5R_{200}$. In low-mass unrelaxed clusters, no significant temperature contrast is observed, with filamentary gas remaining hotter than its surroundings.

Around $3R_{200}$, the temperature drop is both deeper and sharper for massive relaxed clusters compared to low-mass unrelaxed ones. A similar pattern is seen for the temperature bump between $1$ and $2R_{200}$. 
These trends likely reflect the connection between cluster properties and filament characteristics. Massive clusters tend to be linked to more "well-established" filaments, which are 
denser, hotter, and more spatially extended than those around low-mass systems according to \citep{galarraga2020,galarraga2021}. Also, \cite{angelinelli2021} find a positive correlation between filament temperature and host cluster mass, supporting our results. This naturally leads to larger temperature contrasts between filamentary and isotropic gas around massive clusters.
Relaxed clusters, with a more uniform ICM and homogeneous surrounding circumcluster medium, enhance this contrast further, whereas unrelaxed clusters typically feature a more disturbed environment but are connected to a larger number of filaments \citep{Gouin21}. Together, these might explain why the anisotropic temperature is strongest around massive relaxed systems.

\section{Transverse temperature profiles across filaments \label{sect:Tan}}

\begin{figure*}[t]
    \centering
        \includegraphics[width=0.99\textwidth]{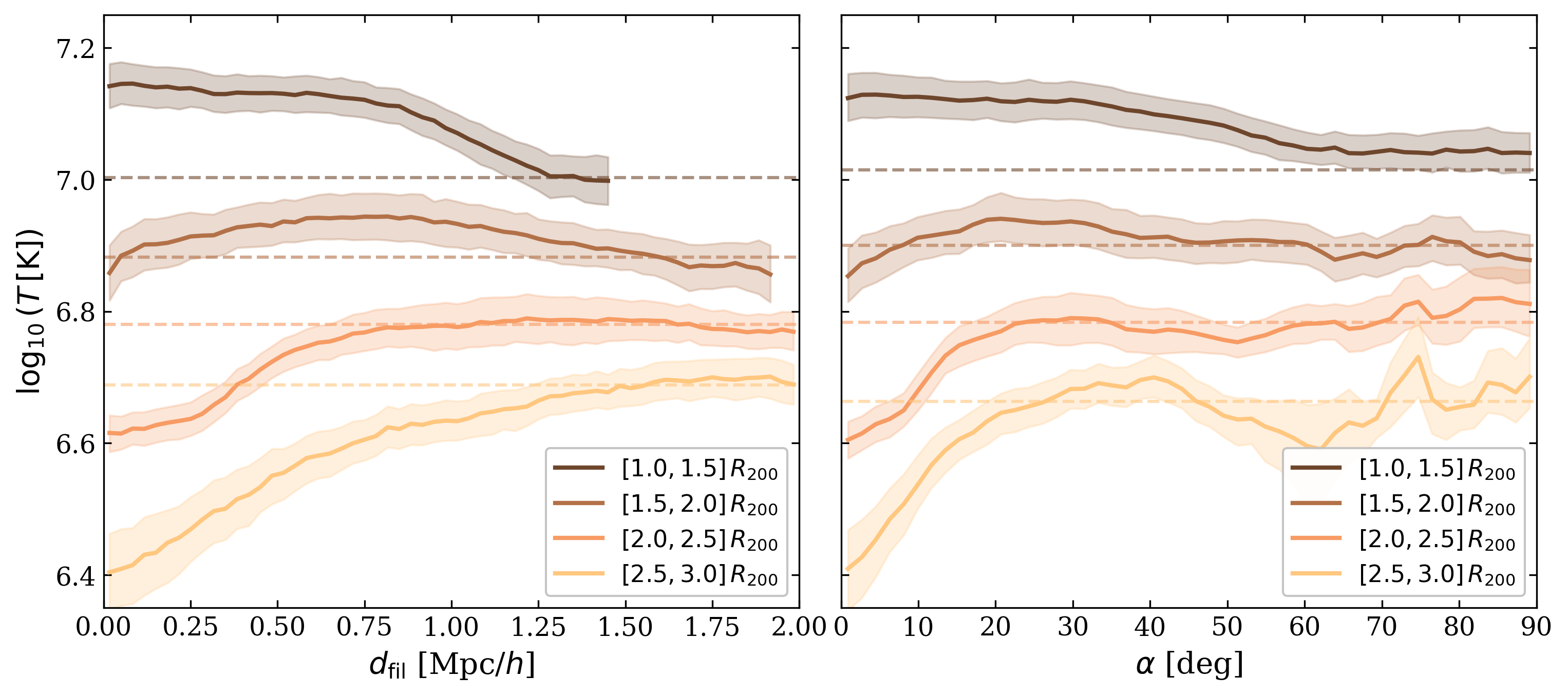}    \caption{Mean temperature profiles as a function of the perpendicular distance from the filament spine $d_{\rm fil}$ (left panel) and the angle to the filament axis $\alpha$ (right panel), for different cluster-centric shells. The uncertainties were estimated via bootstrap resampling, and the dashed lines represent the mean background temperature outside filaments, above $d_{\rm fil}>1.25 {\rm Mpc}.h^{-1}$ for the left panel and $\alpha>60^\circ$ for the right panel.}
        \label{fig:dfil_and_alpha_temp}
\end{figure*}

The top left panel in Fig.~\ref{fig:profile_properties} shows that filaments exhibit both an internal gradient in their properties as a function of $d_{\rm fil}$, and an overall radial gradient toward cluster distance $R$. In contrast to the other thermodynamic quantities (such as density, entropy, and pressure), the temperature displays an inversion of this gradient around $R \sim 2 R_{200}$: at smaller radii, the filament core is hotter than its outskirts, while between 2 to 3.5 $R_{200}$, the trend reverses, with the core becoming colder than the filament edges.
In the following, we further investigated these temperature variations by analyzing the temperature of filament gas in two complementary coordinate systems: radial to the filament spine and angular relative to the filament axis.
In detail, we present here the results for the top 20\% most massive relaxed clusters (see Table \ref{table:MassRange}), as this subsample exhibits the strongest signature of gradient inversion.
Appendix \ref{AnnexeB} shows the results for the low-mass unrelaxed clusters, where no gradient inversion is observed. In these systems, filaments remain consistently hotter than the isotropic background.

\subsection{Radial profile to the filament spine}

The left panel of Fig.~\ref{fig:dfil_and_alpha_temp} shows the radial temperature profile as a function of the perpendicular distance to filaments, for four distinct shells around clusters, from 1 to 3 $R_{200}$.
First, by comparing the absolute temperature values, we confirm a cluster-centric gradient: gas in filaments is hotter near clusters and gradually cools with increasing distance.
Second, we clearly identify an inversion of the temperature gradient perpendicular to filaments, with the transition occurring around $1.5-2\ R_{200}$.
In the vicinity of clusters ($1-1.5R_{200}$, dark brown curve), the temperature decreases slightly but significantly with the angle to the filament axis, suggesting that filaments align with the hottest directions within clusters.
Notice that, in the inversion region, around $1.5-2\ R_{200}$, there is a slight bump in temperature at filament borders.
These behaviors appear consistent with a possible virial shock at the interface between infalling filamentary gas and the ICM. 
In contrast, outside clusters ($2-3\ R_{200}$, orange and light orange curves), the temperature increases with perpendicular distance to filaments. 
For example, between 2.5 and 3 $R_{200}$, the temperature rises from $\sim 10^{6.4}\,\mathrm{K}$ in the filament core to $\sim 10^{6.7}\,\mathrm{K}$ at larger distances, i.e., roughly a factor of 2 increase.

\subsection{Angular profile to the filament axis}

The angular temperature profile is shown in the right panel of Fig.~\ref{fig:dfil_and_alpha_temp}, confirming the radial and transverse trends discussed in Sect. \ref{Physics properties}.
Close to the cluster core, the temperature is slightly higher along the filament axis than in perpendicular directions. Moving toward the outskirts, between 2 and $3R_{200}$, filaments appear colder for angles $\alpha < 20^\circ$, in agreement with the results of \cite{Rost2024}. 
Interestingly, this transition occurs at a nearly constant opening angle ($\alpha \sim 20^\circ$), while the corresponding perpendicular distance increases with radius (from $d_{\rm fil} \sim 0.75 {\rm Mpc}.h^{-1}$ at $2-2.5R_{200}$ to $d_{\rm fil} \sim 1.5{\rm Mpc}.h^{-1}$ at $2.5-3R_{200}$ in the left panel). 
This could indicate that filamentary gas penetrates the circumcluster medium in a conical geometry, with colder gas confined within a widening angular region around the filament axis.

\section{Gas dynamics: Accretion and ejection rate}
\label{Gas dynamics}

In this section, we examine the radial velocity of gas in both isotropic and anisotropic measurements to identify any preferential flow of the cosmic gas around clusters. We also aim to understand the dynamics of the gas by analyzing how it is accreted and ejected in both directions. We first analyze the gas velocity conditional probability to identify preferential inflow and outflow directions before investigating how the gas falls within filaments. We then investigate possible radial velocity gradients with the distance to filaments and their relation with gas temperature.

\subsection{Conditional probability of gas radial velocity}

\begin{figure}
    \centering
    \includegraphics[width=0.95\linewidth]{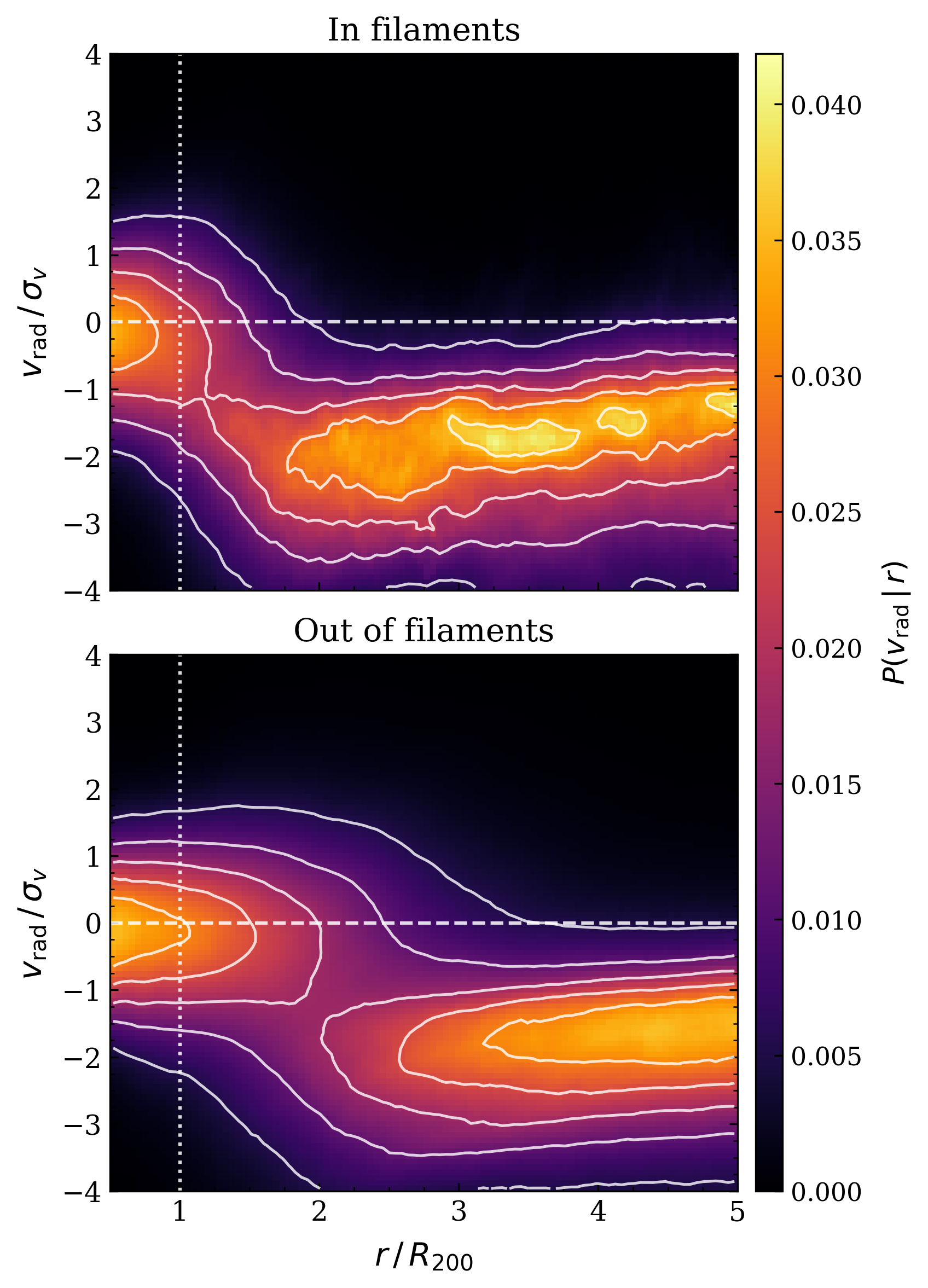}
    \caption{2D histogram of the conditional probability distribution $P(v_{\rm rad} \mid r)$ of radial velocities as a function of cluster-centric distance. Positive (negative) radial velocities indicate outflows from (infall toward) the cluster center. Top panel: Gas cells located within a projected distance $d_{\rm fil} \leq 0.5{\rm Mpc}.h^{-1}$ from the nearest filament axis. Bottom panel: Gas outside filaments ($d_{\rm fil} > 0.5{\rm Mpc}.h^{-1}$). The five white contours represent linearly spaced iso-probability levels between $15\%$ and $85\%$. Results are stacked over the 415 clusters of our sample.}
    \label{fig:vrad2D_histogram}
\end{figure}

To characterize the gas flow from filaments into clusters, we examined the gas radial velocity.  
This quantity was computed by projecting the velocity vector onto the 3D radial direction with respect to the cluster center. The resulting scalar accounts for the direction of motion (inward or outward) via the dot product of the position and velocity vector, and is defined as follows \citep[similar to][]{Oman_2013}: 
\begin{equation}
v_{\text{rad}}^{(i)} =
\left(\vec{v}_{\text{gas}}^{(i)} - \vec{v}_{\text{cluster}}\right)
\cdot
\frac{\left(\vec{x}_{\text{gas}}^{(i)} - \vec{x}_{\text{cluster}}\right)}
{\left\|\vec{x}_{\text{gas}}^{(i)} - \vec{x}_{\text{cluster}}\right\|},
\end{equation}
where $\vec{v}_{\text{rad}}^{(i)}$ is the radial velocity of the i-th gas element, $\vec{v}_{\text{gas}}^{(i)}$ and $\vec{x}_{\text{gas}}^{(i)}$ are respectively its velocity and position vectors, and $\vec{v}_{\text{cluster}}$ and $\vec{x}_{\text{cluster}}$ denote the bulk velocity and position of the cluster center. The radial direction is defined as the unit vector pointing from the cluster center to the gas element
i.e., $(\vec{x}_{\text{gas}}^{(i)} - \vec{x}_{\text{cluster}}) / \|\vec{x}_{\text{gas}}^{(i)} - \vec{x}_{\text{cluster}}\|$. A negative sign indicates gas accretion toward the cluster, and a positive sign indicates gas ejection. 
Notice that, for each cluster, the radial gas velocity is normalized by a characteristic velocity scale defined as the standard deviation of the radial gas velocities within $R_{200}$, $\sigma_v = \mathrm{std}(v_{\mathrm{rad}})$, representing the internal velocity dispersion of the ICM.

Based on this definition, Fig.~\ref{fig:vrad2D_histogram} shows the 2D histogram of normalized radial velocity  $v_{\rm rad}/ \sigma_v$ versus normalized cluster-centric distance $r/R_{200}$, averaged over the 415 clusters.
Moreover, this histogram uses equal-width linear bins along both axes, with each bin representing the number of particles whose velocities and distances fall within $[v_i, v_{i+1}) × [r_j, r_{j+1})$.
The conditional probability distribution of the radial velocity of the gas is written as follows: 
\begin{equation}
P\!\left(v_{\rm rad} \mid r\right) = \frac{N(r, v_{\rm rad})} {\sum\limits_{v_{\rm rad}} N(r, v_{\rm rad})}.
\end{equation}
With $P(v_{\rm rad} \mid r)$ normalized such that $\int P(v_{\rm rad} \mid r)\,\mathrm{d}v_{\rm rad} = 1$ in each radial bin.
Consequently, Fig. \ref{fig:vrad2D_histogram} illustrates the probability of a particle possessing a specific velocity at a given distance from the cluster center.

The velocity–position diagrams in Fig.~\ref{fig:vrad2D_histogram} highlight the distinct accretion patterns of gas inside and outside filaments. 
Beyond $3R_{200}$, warm gas predominantly accretes along filaments, though a nonnegligible fraction of infall occurs outside filaments. 
Between $1$ and $3 R_{200}$, velocities are bimodal: high-velocity inflows occur mainly along filaments, while lower-velocity outflows are almost only outside filaments. Indeed, because we adopted a constant $d_\mathrm{fil}=0.5{\rm Mpc}.h^{-1}$, we observe a very small fraction of ejected gas with $v_\mathrm{rad} > 0$ located near clusters within filaments.
Within clusters ($r < R_{200}$), inflows and outflows are roughly symmetric, consistent with quasi-hydrostatic equilibrium.

This accretion process, illustrated in Fig. \ref{fig:vrad2D_histogram}, shows that gas moving toward clusters is primarily slightly accelerated within filaments, but is also present outside them. This can be explained by two factors. First, given that we adopted a constant $d_\mathrm{fil}=0.5{\rm Mpc}.h^{-1}$, the volume ratio of gas outside to inside filaments increases rapidly with cluster-centric distance. Second, the motion of gas outside filaments likely reflects the large-scale flow of material moving simultaneously toward both filaments and cluster center. 
These velocity diagrams show the same typical accretion behavior observed for galaxies: fast-infalling galaxies approach clusters with increasing velocity, followed by backsplash ejection, before eventually returning toward equilibrium \citep[see, e.g.,][]{ArthurRam,Mostoghiu2019}. Focusing on gas accretion, our findings—consistent with \cite{Rost2021}—indicate that fast-moving gas is primarily located within filaments, whereas most of the ejected gas lies outside filaments, mainly between $1$ and $3R_{200}$. In addition, some low-velocity gas may simultaneously flow toward both filaments and clusters. This suggests that gas accretion onto clusters occurs through coherent filamentary inflows and more diffuse isotropic infall from the surrounding medium. Gas within filaments follows preferential, dynamically organized pathways driven by the large-scale structure, while gas located between filaments is still gravitationally attracted toward the cluster, in a less structured way \citep{Lebeau2025,Zinger_2018,Lu2024}.

\subsection{Distribution of radial velocity inside filaments}
\begin{figure}
    \centering
    \includegraphics[width=1\linewidth]{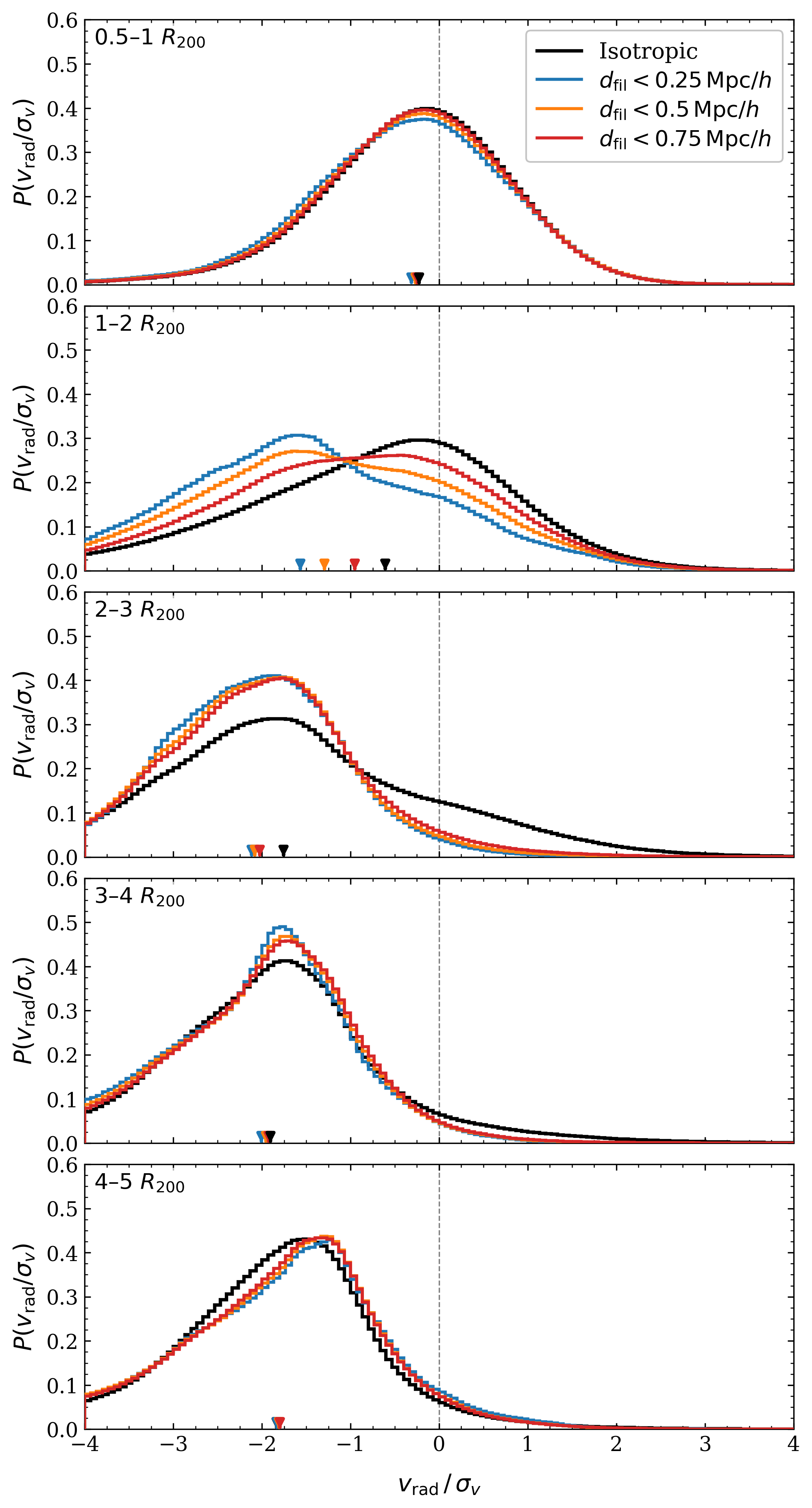}
    \caption{Normalized 1D histogram of the radial velocities for different shell of distance from the cluster center: $[0.5,1],\ [1,2],\ [2,3],\ [3,4],\ [4,5]$, in $R_{200}$ unit, with the isotropic gas case in black and the anisotropic cases in blue, orange, and red for $d_{fil}=[0.25;0.5;0.75]{\rm Mpc}.h^{-1}$, respectively. The colored arrows represent the median values of the respective histograms.}
    \label{fig:vrad1D_in_out_fil}
\end{figure}
\begin{figure}[h]
    \centering
    \includegraphics[width=1\linewidth]{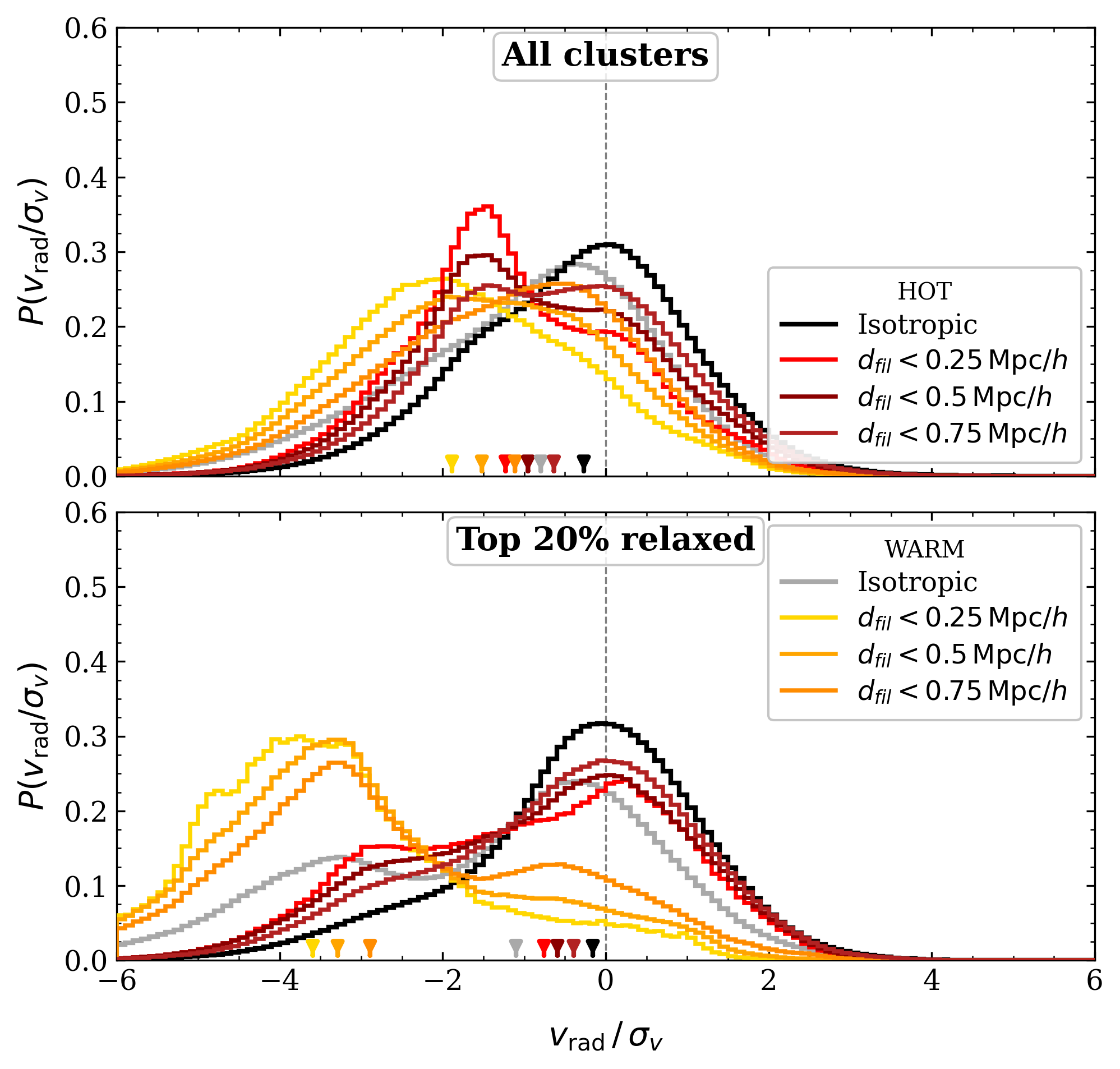}
    \caption{Normalized radial velocity distributions $P(v_{\rm rad}/\sigma_v)$ 
    in the $[1,2]R_{200}$ shell for two cluster samples. 
    \textit{Top panel:} All 415 clusters. 
    \textit{Bottom panel:} Top 20\% most massive relaxed clusters. 
    In each panel, the black and gray curves show the isotropic distribution 
    for hot and warm gas, respectively, while colored curves show the 
    in-filament distributions for $d_{\rm fil} < 0.25,\,0.5,\,0.75\,{\rm Mpc}.h^{-1}$. 
    (yellow to dark orange for warm gas; red to dark red for hot gas).}
    \label{fig:vrad_histo_relaxed_WARM_HOT}
\end{figure}

To complete this analysis, we decomposed the radial velocity distribution into five radial shells centered on clusters and distinguished between gas located inside filaments and the isotropic component (Fig.~\ref{fig:vrad1D_in_out_fil}).
Consistent with previous findings, the first shell ($0.5R_{200} < r < R_{200}$) shows a nearly balanced ratio between inflows and outflows, indicative of the dynamical equilibrium of clusters. 
At the cluster outskirts ($1R_{200} < r < 2R_{200}$), gas falls faster in filaments, while low-velocity infall dominates outside filaments; outflows remain mostly confined to non-filamentary regions. The highest infall velocities are observed in filament centers ($d_{\rm fil}=0.25{\rm Mpc}.h^{-1}$), decreasing toward the filament edges.
From 3 to 2$R_{200}$, the anisotropic accretion through filaments dominates, with negligible outflows. Beyond  $3R_{200}$, outflows are minimal, and infall velocities converge between filamentary and non-filamentary regions.

To further investigate the $[1,2]R_{200}$ region, Fig.~\ref{fig:vrad_histo_relaxed_WARM_HOT} shows the probability distribution function of $P(v_{\rm rad}/\sigma_v)$ for the full sample of 415 clusters and for the $20\%$ most massive clusters in the relaxed subsample.
For the full sample, warm and hot gas exhibit similar velocity distributions, indicating no clear dynamical difference between inflowing and outflowing material. In contrast, the massive relaxed subsample shows a clear separation: warm gas is predominantly associated with fast inflows along filaments, while outflows are dominated by hot gas.
This behavior is consistent with the enhanced thermodynamic contrast observed in Fig.~\ref{fig:ratio_relaxation_mass}, where massive relaxed systems show up to $\sim40\%$ differences between filament and outskirts temperatures, reflecting a distinction between accreted and ejected gas.

Finally, in Fig. \ref{fig:vrad_mach} left panel, we present a 2D map of the radial velocity field $v_{\rm rad}$ for a representative cluster, projected along the XY plane. The color map encodes the sign and magnitude of the radial velocity, and the overlaid velocity vectors highlight the full kinematic structure of the gas around the cluster.
The map clearly shows that the bulk of the gas within $\sim 3\,R_{200}$ is dominated by infall, as expected from the previous results. However, the central region exhibits a mixture of positive and negative radial velocities, reflecting the gas infalling from filamentary structures and gas that has been partially expelled by the cluster. We note that the gas motion is not purely radial, as significant tangential components are present, especially in the inner regions. These tangential motions drive part of the gas expelled between $1$ and $2\,R_{200}$ back into the filaments, where it falls back and is reaccreted into clusters.

\section{Discussion\label{sect:Discussion}}

We summarize here our findings and compare them to past studies on clusters and filaments using hydrodynamical simulations, with the aim of describing a coherent gas accretion picture from filaments to clusters. Secondly, we open the discussion on shock mechanisms. 

\subsection{A coherent gas accretion picture \label{sect:litt}}

\begin{figure*}[!t]
    \centering
    \begin{minipage}[t]{0.49\textwidth}
        \vspace{0pt}
        \centering
        \includegraphics[width=\textwidth]{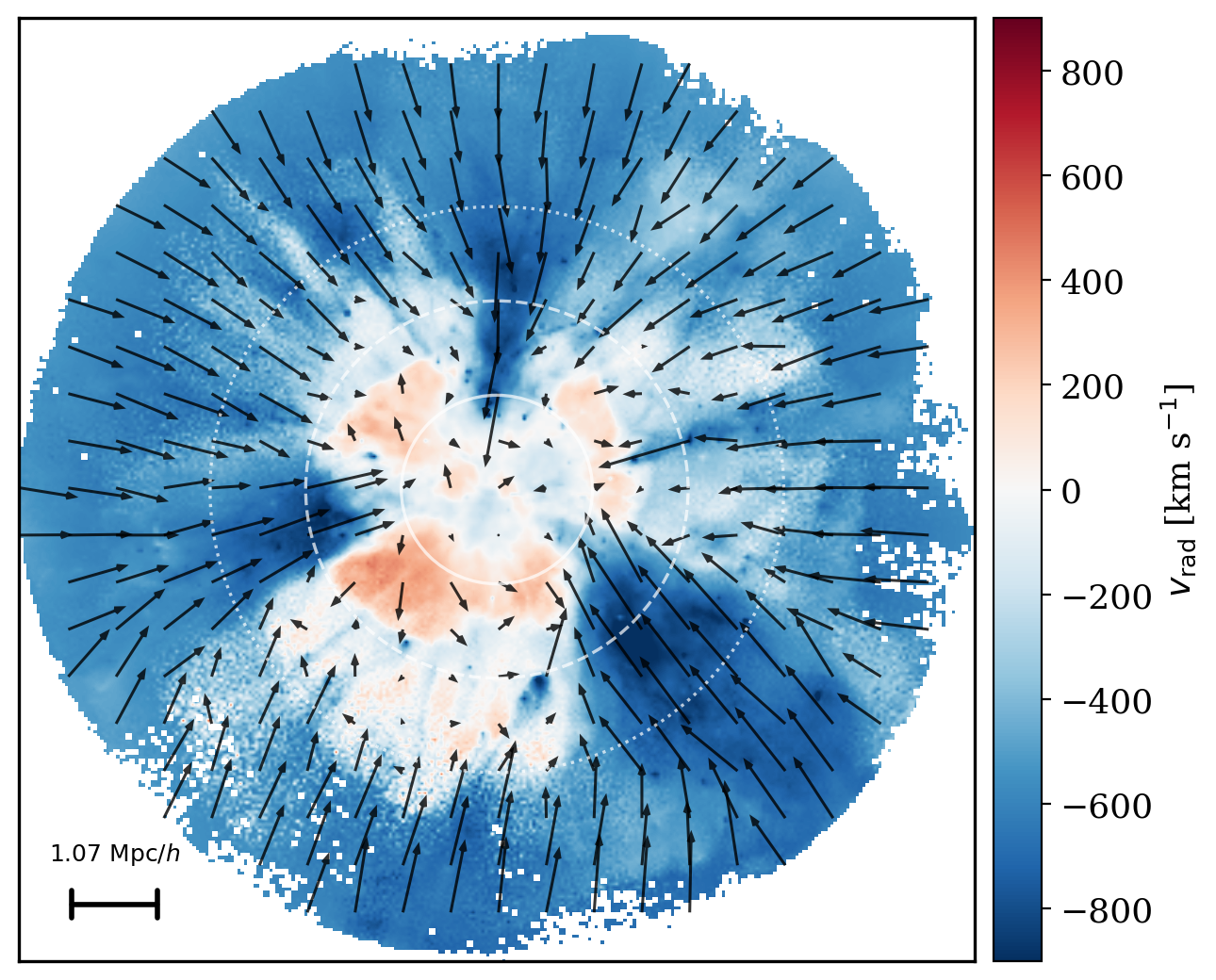}
    \end{minipage}
    \hfill
    \begin{minipage}[t]{0.49\textwidth}
        \vspace{0pt}
        \centering
        \includegraphics[width=\textwidth]{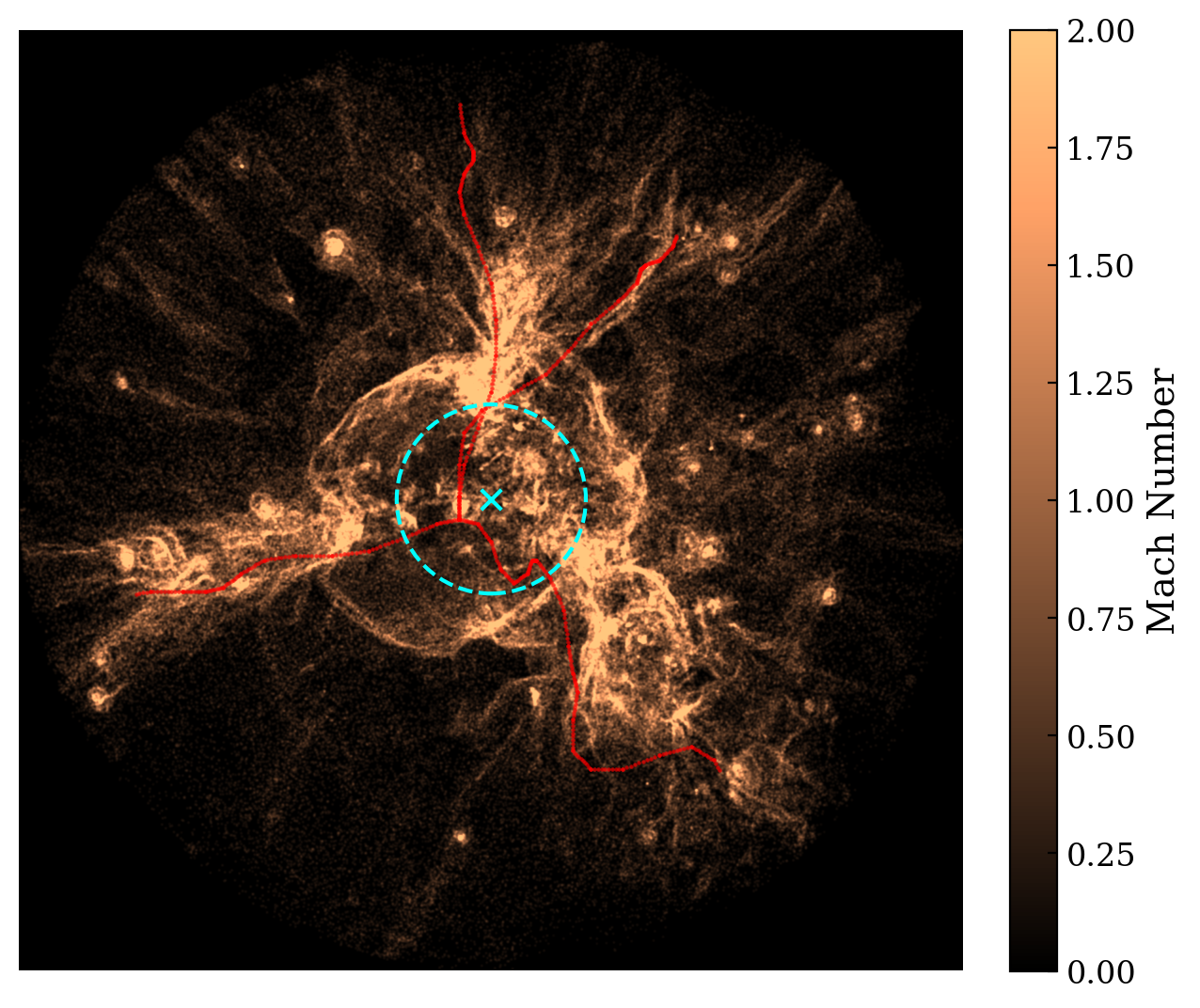}
    \end{minipage}
    \caption{
    \textit{Left panel:} Radial velocity map of a given cluster, where the blue and red regions represent infalling and ejected gas, respectively. The white circles mark 1, 2, and 3 $\,R_{200}$, respectively. The arrow lengths are proportional to the velocities.
    \textit{Right panel:} Mach number map for the same cluster with red lines indicating the projected filamentary skeleton.
     \label{fig:vrad_mach}}
\end{figure*}

An increasing number of studies have recently explored baryonic physics in the cosmic web using simulations, particularly focusing on the complex transition from diffuse gas in filaments to the hot virialized gas in galaxy clusters. Using the \TTH project, previous works have investigated various processes and properties in these regions, including thermodynamics \citep{Rost2024}, velocity fields \citep{Rost2021}, diffuse gas anisotropies \citep{Lokken2023}, and galaxy quenching \citep{Kotecha2022}.

Specifically, consistent with our findings, \cite{Rost2024} identify an elevated temperature and entropy outside the filaments, from approximately $2$ to $4R_{200}$. While the overall trends in filament and cluster temperature evolution are similar, absolute values and radial scales differ from our results. For instance, while \cite{Kotecha2022} reports filaments penetrating deep into clusters with a cold core extending down to $0.3\,R_{\mathrm{vir}}$, our results suggest that filament cores are already significantly heated at $\sim 1.5\,R_{200}$. 

This radial limit is, however, consistent with \cite{Vurm2023}, who report similar trends in the gas thermodynamics of in and out of filaments around a single massive cluster in the Cluster-EAGLE simulation \citep{bahe2017}.

Based on our results and in line with recent studies, we propose the following scenario for gas accretion from cosmic filaments onto galaxy clusters, as illustrated in Fig.~\ref{fig:Picture}. At large cluster-centric distances ($\gtrsim 4R_{200}$), the warm gas is predominantly confined within filamentary structures and is infalling toward the clusters, as shown in Fig. \ref{fig:profile_properties}. Closer to cluster environments, around $3.5R_{200}$, the ambient medium reaches temperatures comparable to those within the filaments. 
In these regions, from $3.5$ to $2R_{200}$, the filament's gas thermalization is not efficient due to the higher infall velocities (see Fig.~\ref{fig:vrad2D_histogram} and Fig.~\ref{fig:vrad1D_in_out_fil}), resulting in significantly colder filaments than in the isotropic direction, in particular around massive clusters (see Fig.~\ref{fig:ratio_relaxation_mass}). However, we observe a tangential temperature gradient, increasing with perpendicular distance to filaments (see Fig.~\ref{fig:dfil_and_alpha_temp}), showing the beginning of filament thermalization at these borders.
Finally, at cluster outskirts, from $1$ to $2R_{200}$, fast infalling gas in filaments encounters the hot and dense ICM. 
At these radii, gas in filaments exhibits a tangential velocity gradient, showing that gas in filament cores is faster than at filament borders, as shown in Fig.~\ref{fig:vrad1D_in_out_fil}. In particular, for massive relaxed clusters, we observe a clear bimodality between fast warm gas infalling from filament cores and hot gas slowly ejecting isotropically (see Fig.~\ref{fig:vrad_histo_relaxed_WARM_HOT}).

\subsection{Shock signatures in the cluster outskirts \label{sect:shocks}}

As described by \citet{Molnar2009}, two principal categories of shocks are commonly identified in galaxy cluster environments: virial shocks, which occur near the cluster's virial radius with Mach numbers of $\mathcal{M}\sim2$ \citep{Vazza2011}, and accretion shocks at several virial radii with higher Mach numbers $\mathcal{M}\sim100$ \citep{Baxter2021}.
We focus on our two regions of interest, at cluster outskirts and in the fast infalling regions, to discuss these potential shocks in the framework of our findings. 

First, between $1$ and $2R_{200}$, the entropy outside filaments increases from $\sim 0\%$ up to $\sim 50\%$ higher than inside filaments. This sharp rise likely reflects heating or mixing driven by shocks or turbulence. This feature is consistent with the presence of virial shocks, expected near the cluster boundary ($\sim 0.9$–$1.3R_{200}$; \citealt{Molnar2009,ZangC2020}).
To investigate this virial shock, we plot in the right panel of Fig.~\ref{fig:vrad_mach}, the Mach number distribution around a given relaxed and massive cluster. As shown here and tested over our cluster sample, we clearly observe a shock front near $R_{200}$ with $\mathcal{M} \sim 2$ in most of the clusters.

\begin{figure}
    \centering
    \includegraphics[width=0.49\textwidth]{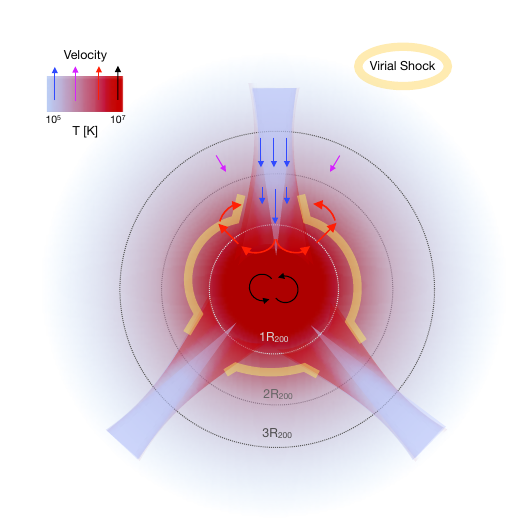}
    \caption{Schematic illustration of the global accretion picture \label{fig:Picture}}
\end{figure}

This supports our gas accretion scenario, in which fast infalling gas along filaments shocks against the ICM, leading to enhanced temperature anisotropy and mixing between warm infalling and hot ejected gas, as illustrated in Fig. ~\ref{fig:Picture}.
We remind readers that only relaxed clusters show a clear heating phase within filaments around $R_{200}$, consistent with \cite{Baxter2021}. Indeed, they find that the virial shock radius depends on the cluster dynamical state, with relaxed clusters exhibiting shocks at larger radii.

Second, at larger distances from clusters (from $\sim 2$ to $\sim 4 R_{200}$), we find strong variations in both entropy and temperature in filaments compared to the isotropic average. However, we do not observe high Mach numbers, as expected from \cite{Molnar2009} for accretion shock. 
One possible explanation is that the gas in the cluster outskirts at $z=0$ is already significantly preheated, which reduces the contrast across shocks and leads to lower effective Mach numbers. This is consistent with previous studies showing that feedback processes can modify the thermodynamic state of the gas and weaken shock signatures in cluster environments, especially at $z=0$ \citep{Schaal_2016}.
In contrast, we tend to observe weak shock signatures around filaments, as illustrated in the right panel of Fig.~\ref{fig:vrad_mach}, consistent with recent studies of small-scale filament shocks at higher redshift \citep{Ramsoy_2021,Lu2024,Yao2025,Medlock2026}.
A more detailed investigation of these features will require further work and is beyond the scope of this paper; we defer a comprehensive analysis to a future study.

\section{Conclusion\label{Conclusion}}

This study aimed to physically characterize cosmic gas accretion and its impact on thermodynamic properties.
We focused on the transition regions between cosmic filaments and galaxy clusters, where structure assembly meets multiphase gas mixing, from warm cosmic-web gas to hot cluster cores.
In practice, we analyzed a statistical sample of 415 groups and clusters of galaxies from the IllustrisTNG simulation at $z=0$.
Our results highlight an accretion scenario capturing the complex interplay between infalling filamentary gas and the cluster potential, as summarized below.

First, outside cluster environments ($r>4R_{200}$), warm gas is funneled along filaments toward clusters, exhibiting relatively uniform velocities across the filament width. 
Then, below $\sim3.5-4R_{200}$, filaments begin to penetrate the extended warm environment surrounding clusters.
In this infalling region, from 4 to 2 $R_{200}$, a clear transverse gradient in temperature and entropy is observed, with higher values toward the filament outskirts.
This likely results from frictional heating at the filament boundaries with the circumcluster medium, while gas along the filament spine remains comparatively colder and continues to flow inward at higher velocities. Geometrically, we find that this anisotropic inflow of colder gas tends to form a conical structure.

Second, as filamentary gas reaches the cluster outskirts at $\sim 1-2 R_{200}$, the infalling gas starts to collide with the ICM, which can generate a virial shock that heats the anisotropically infalling gas.
In detail, we observe two distinct gas behaviors in filaments: 
(i) fast infalling warm gas in the filament cores penetrates deeper into clusters; and
(ii) at the filament outskirts, gas is decelerated, encountering hot gas ejection from the ICM.
Indeed, the gas does not immediately settle within the clusters and is instead partially pushed back into the cluster outskirts and out-to-filaments at lower velocities. 
This cyclical process of accretion and partial ejection contributes to increasing gas entropy and temperature in cluster outskirts.

Finally, we find that this global accretion picture and its associated anisotropic signatures vary with cluster properties: massive and dynamically relaxed clusters exhibit the strongest imprints.
Our results suggest that such halos have a well-established filament–cluster connection in the absence of recent major mergers, leading to efficient gas cooling in filaments and shock heating at the filament–ICM interfaces.

To conclude, our results demonstrate that galaxy clusters cannot be described as isotropic systems, as filamentary accretion imprints a strong directional dependence on gas properties, shaping their thermodynamic structure out to several virial radii. These anisotropic signatures could have direct observational consequences, particularly for X-ray and Sunyaev–Zel’dovich measurements. In this context, X-ray cluster observing programs including CHEX-MATE \citep{CHEXMATE}, as well as those conducted with X-ray spectroscopy missions such as XRISM \citep{xrism2020} and New-Athena \citep{Cruise_2024}, will provide unique opportunities to probe the imprints of cosmic gas accretion around clusters.

\begin{acknowledgements}

We thank the IllustrisTNG collaboration for public access to the data used in this work. We also thank Clotilde Laigle, Maelie Mondelin, and Charlotte Welker for useful discussions. We thank the anonymous referee for the constructive feedback.

\end{acknowledgements}

\bibliography{max}
\bibliographystyle{aa}

\begin{appendix}
\section{Filament finder methodology \label{appendixA}}

\begin{figure}
    \centering
    \includegraphics[width=0.85\linewidth]{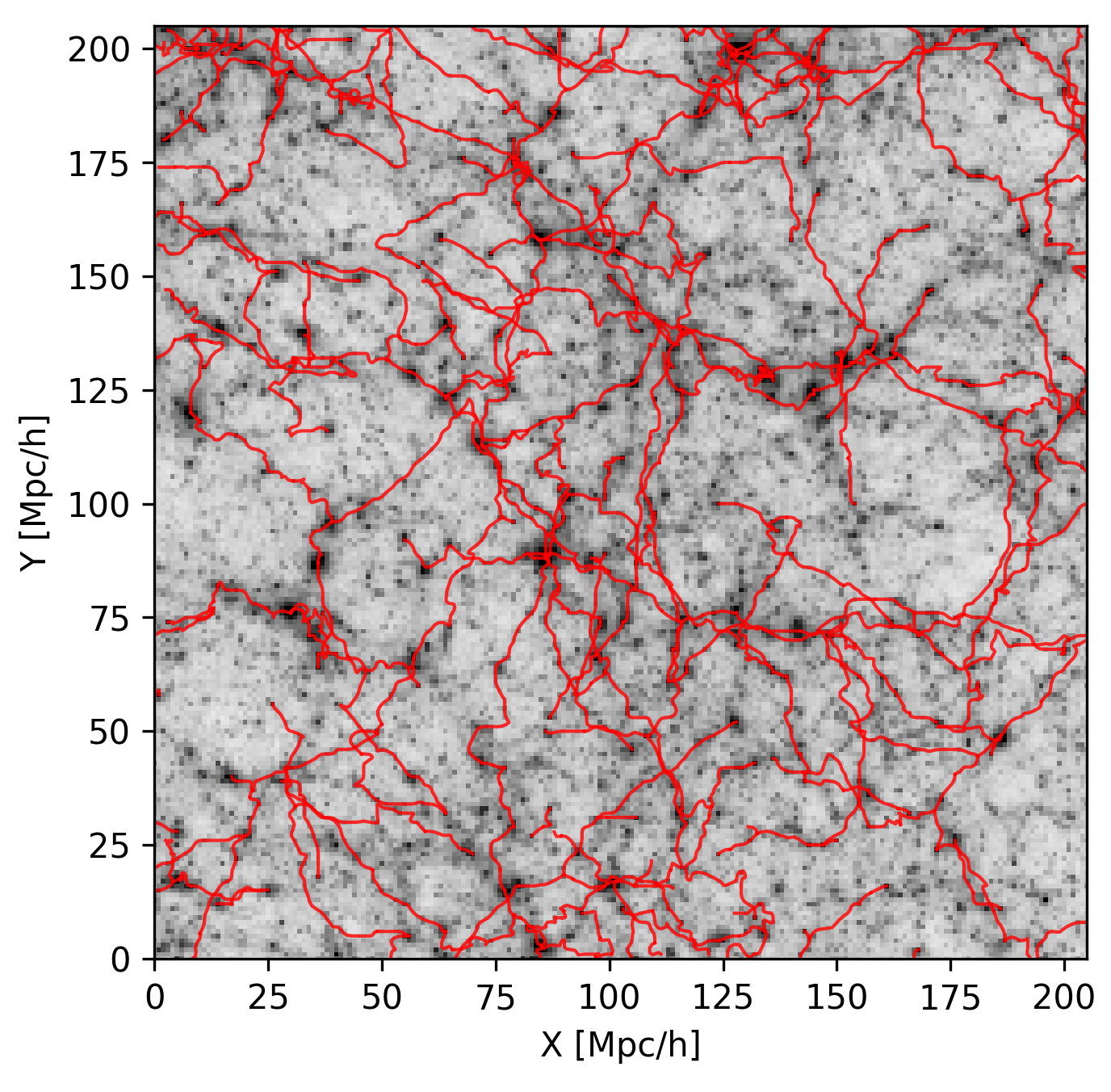}  
    \includegraphics[width=0.85\linewidth]{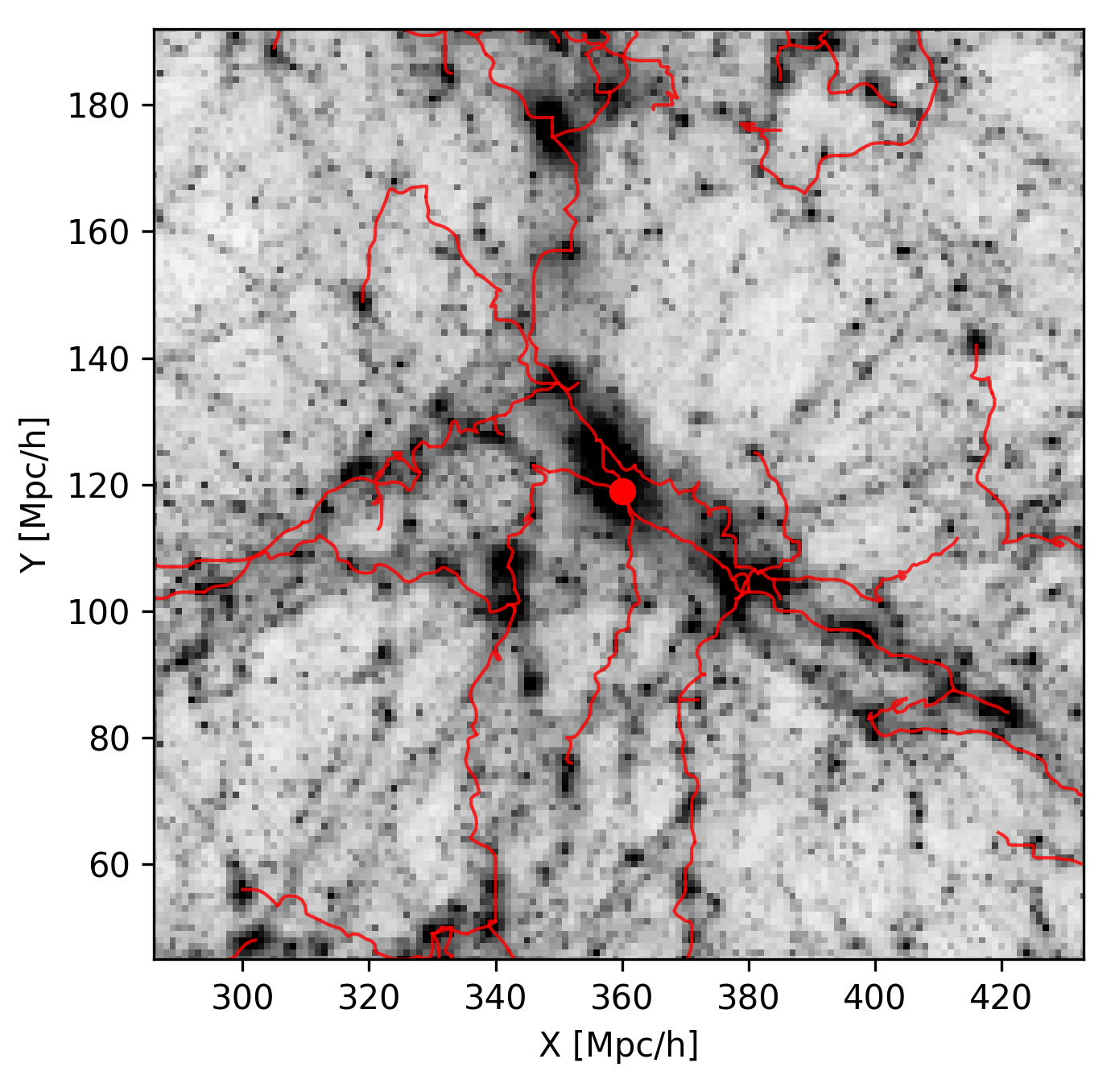}
    \includegraphics[width=0.75\linewidth]{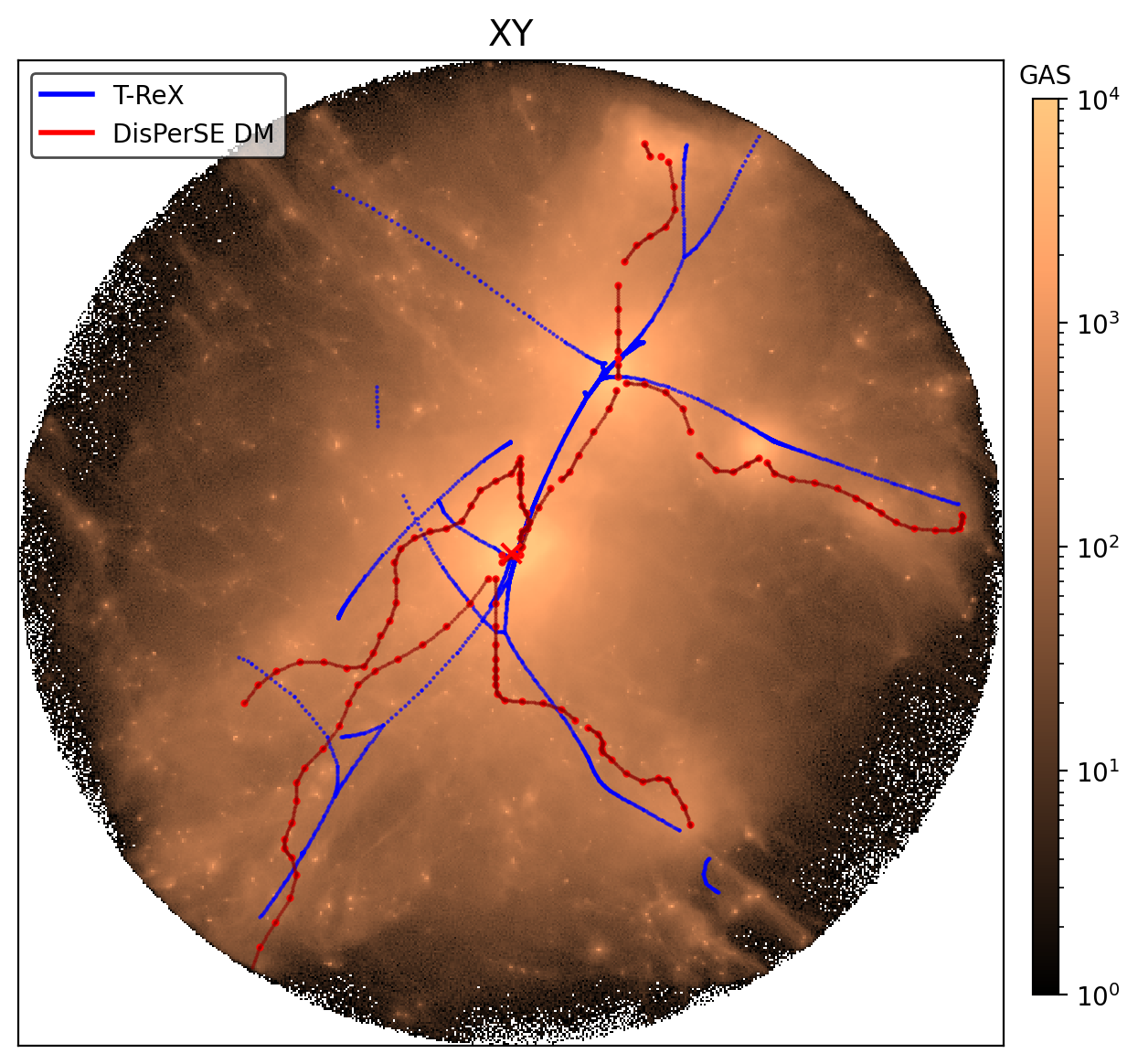}
    \caption{Representation of the filaments found by DisPerSE at large and small scales (top and middle panels; on the DM distribution) and their comparison to T-ReX filaments (bottom panel). Note that the T-ReX algorithm is computed on the galaxy distribution, and DisPerSE on the DM one.}
    \label{fig:TREX_Disperse_appendix}
\end{figure}

In this section, we detail our methodology for detecting the cosmic web skeleton with the DisPerSE algorithm on the DM distribution, and then we compare it to T-ReX filaments extracted from the galaxy distribution. This comparison allows us to assess the robustness of filament identification across different tracers and to quantify potential biases between the DM skeleton and its observational counterpart traced by galaxies.

First, we chose to compute the cosmic web skeleton from the DM distribution rather than gas, as the latter is subjected to processes such as accretion, ejection, and shocks, which can bias filament patterns in these critical regions.
As explained in Sect.~\ref{sect:disperse}, we applied the DisPerSE algorithm to the smoothed 3D DM density field, adopting an absolute persistence threshold of $T = 6000$ and a smoothing parameter of $S = 1$. These choices are motivated by the extensive analysis of \cite{Bahe2025}. In detail, after testing different DisPerSE parameterizations, we chose these values as a good compromise between large-scale filament detection connected to clusters and avoiding small bridges between substructures.
As we can see in the top and middle panels of Fig. ~\ref{fig:TREX_Disperse_appendix}, the resulting filament skeleton consistently follows the underlying DM distribution. 

Secondly, the Tree-based Ridge eXtractor filament finder \citep[T-ReX,][]{trex2020, trex2021} is an alternative filament detection method applied to assess the robustness of our skeleton obtained with DisPerSE.  It relies on a minimum spanning tree technique, i.e., a graph algorithm, and on the gaussian mixture model.
As a result, T-ReX is particularly efficient for processing discrete point sets, such as galaxies, but is less suitable for continuous fields, such as DM or gas, in the context of simulations.
Therefore, we apply the T-ReX algorithm on galaxies with a stellar mass cut of $M_{*} > 10^9M_{\odot}$ (similar to observational selection).
By assuming a standard parameterization of T-ReX \citep[as used by][]{Gouin21,Gouin25}, we detect large-scale cosmic filaments connected to galaxy clusters.

The Fig.~\ref{fig:TREX_Disperse_appendix} illustrates the filaments identified by T-ReX in blue and DisPerSE in red, for the same massive cluster. As observed in this figure, both tracers identify similar filament patterns, and our analysis yields consistent trends in filament gas properties between the two methods, with radial profiles in good overall agreement.
However, we notice hotter filament cores in the case of T-ReX.
Since its filaments are traced on galaxy distribution, their spines tend to be centered on warm galactic medium, thereby biasing the inferred temperature profiles.
Therefore, we chose to present only our analyses based on DisPerSE DM filaments.

\section{Gas temperature in the case of unrelaxed and non-massive clusters and groups \label{AnnexeB}}
Fig.~\ref{fig:temp_unrelaxed_unmassive} illustrates the temperature as a function of $d_{\rm fil}$ (top panel) and the angle $\alpha$ (bottom panel) for the 20\% least massive unrelaxed clusters, complementary to Fig.~\ref{fig:dfil_and_alpha_temp}. 
In contrast to the massive relaxed case, the gradient toward the filament spine ($d_{\rm fil} \rightarrow 0$ and $\alpha \rightarrow 0$) is inverse. Indeed, in this case, filament cores are hotter than their surroundings.
\begin{center}
    \includegraphics[width=0.95\linewidth]{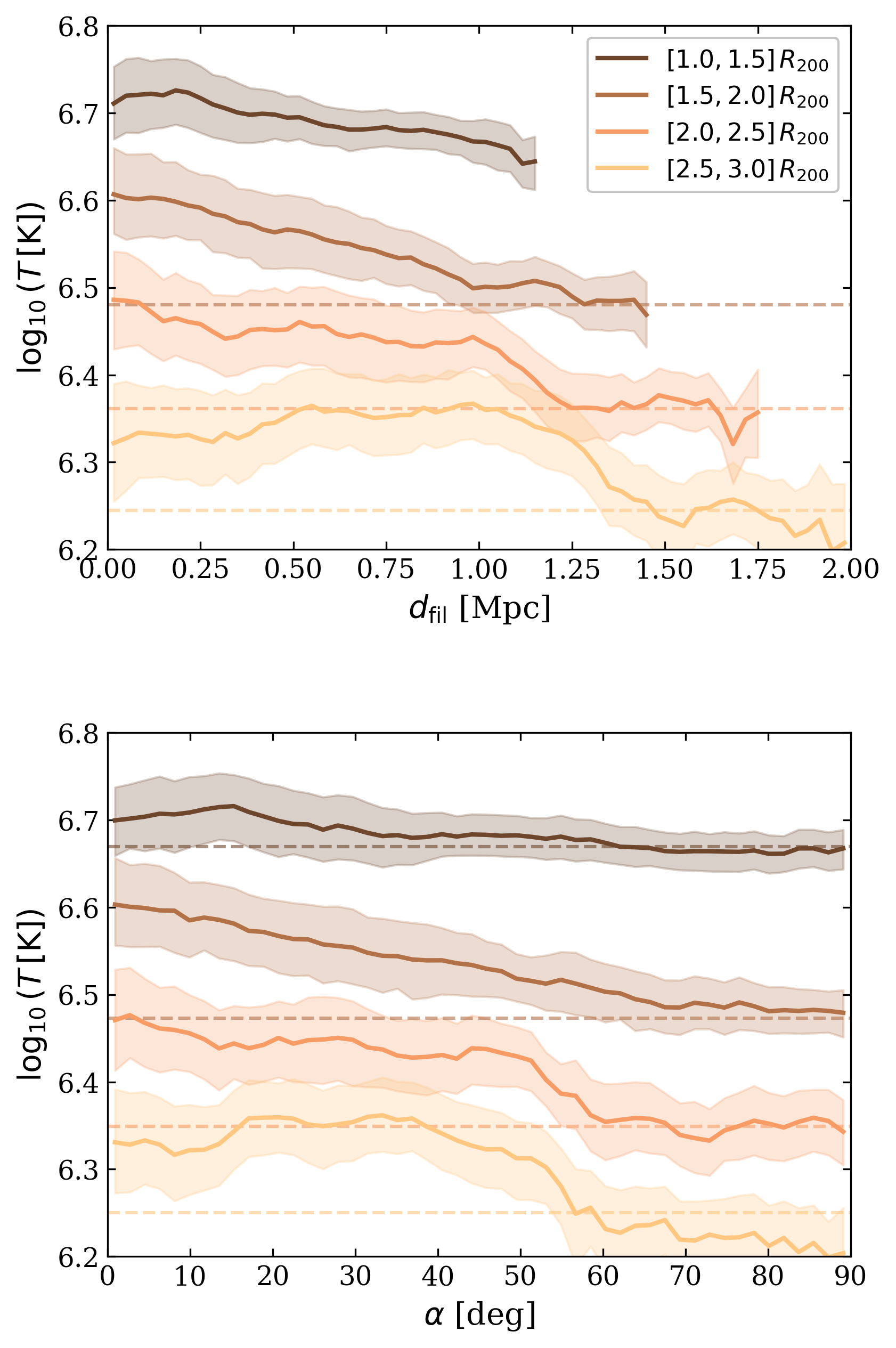}
    
    \captionof{figure}{Same as Fig.~\ref{fig:dfil_and_alpha_temp} but for 20\% least massive clusters of the unrelaxed cluster sample.}
    \label{fig:temp_unrelaxed_unmassive}
\end{center}

\section{Mass-weighted thermodynamic profiles\label{AnnexeC}}

\begin{figure}
    \centering
    \includegraphics[width=0.95\linewidth]{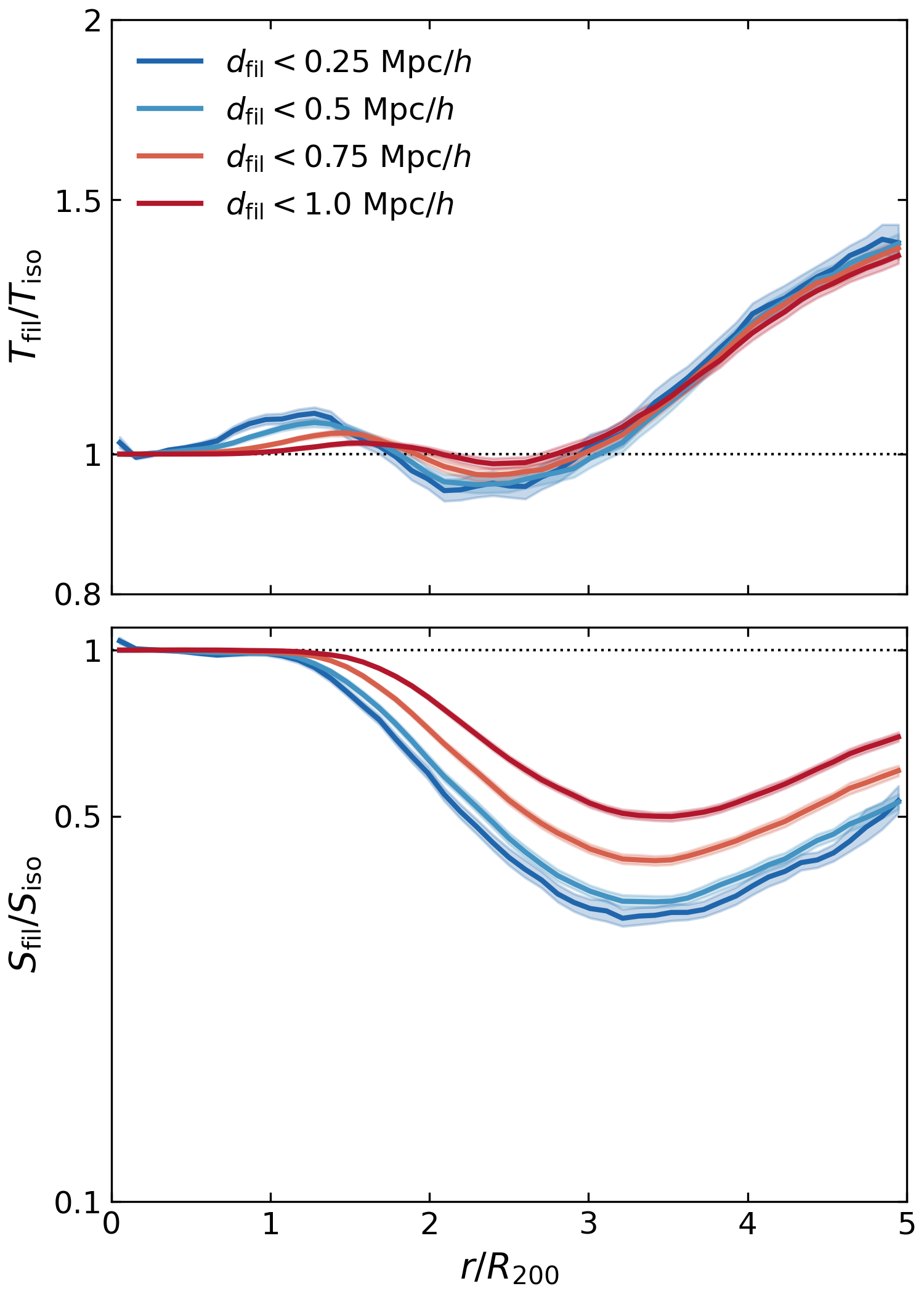}
    \caption{Mass-weighted ratio between the isotropic and anisotropic temperature (top) and entropy (bottom) as function of cluster distance for different filament thickness. }
    \label{fig:ratio_temperature_entropy_mass_weight}
    
    \vspace{1em}
    
    \includegraphics[width=0.95\linewidth]{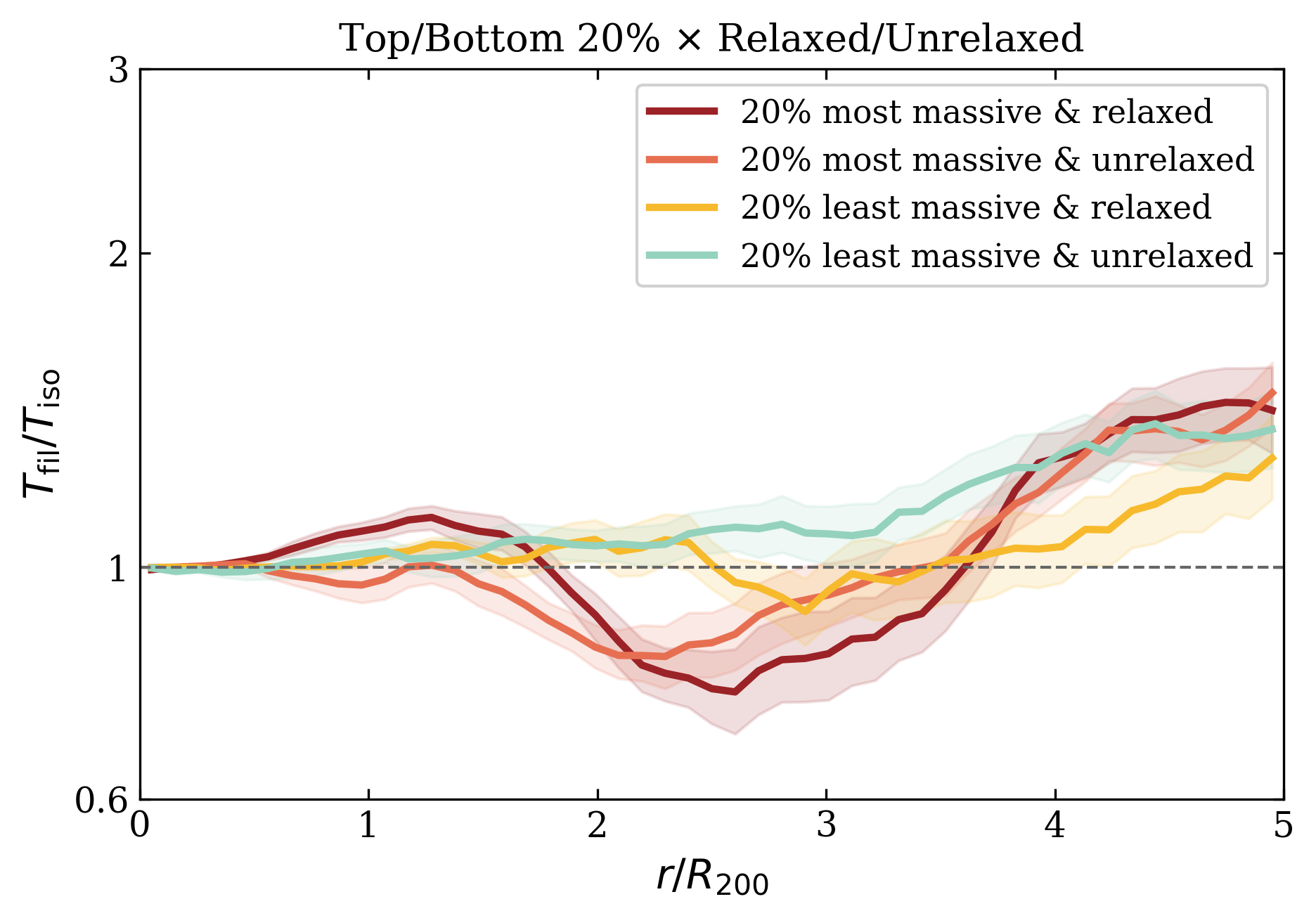}
    \caption{Same as in the bottom panel of Fig.~\ref{fig:ratio_relaxation_mass} but using mass weighting instead of volume weighting.}
    \label{fig:ratiosubsamples}
\end{figure}

Volume weighting was adopted throughout this work to account for the varying volumes of the Voronoi gas cells in the \texttt{AREPO} grid. In moving-mesh hydrodynamical simulations, variations in gas-cell volume arise naturally from the mass-based refinement criterion. Consequently, different weighting schemes can be considered, most notably mass weighting and volume weighting. Whereas mass weighting emphasizes the contribution of the most massive gas cells, volume weighting assigns equal weight to each unit of physical volume.
Since our analysis focuses on characterizing the isotropic-to-anisotropic contrast in diffuse gas, volume weighting is particularly well suited to our purpose. By placing greater emphasis on the most massive cells, mass weighting may preferentially trace denser regions and potentially reduce the relative contribution of the diffuse component, thereby affecting the amplitude of the measured anisotropic signal.
As shown in Fig.~\ref{fig:ratio_temperature_entropy_mass_weight}, the amplitude of the temperature ratio is less pronounced when mass weighting is adopted, as expected. Nevertheless, the overall trends remain nonnegligible. This is further confirmed for the same subsamples considered in Fig.~\ref{fig:ratio_relaxation_mass}, as shown in Fig.~\ref{fig:ratiosubsamples}.
For a detailed investigation of the impact of different weighting schemes, we refer the reader to the section ~4.2 in \citet{braspenning2024}, who examined the effects of mass, volume, and X-ray weighting on the thermodynamic profiles of massive galaxy clusters in the \texttt{FLAMINGO} simulations. They found that, compared with volume weighting, mass weighting tends to exhibit lower temperatures, lower entropies, and higher metallicities, particularly within cluster cores ($\lesssim 0.1$~Mpc). However, the differences induced by the choice of weighting scheme remain within the scatter among different observational datasets.

\end{appendix}
\end{document}